\documentclass[aps,prc,twocolumn,groupedaddress]{revtex4-2}
\usepackage{amsmath}
\usepackage{graphicx}
\usepackage{color}
\begin{document}
%\linenumbers

%\preprint{}

\title{Continuum-discretized coupled-channel calculations for $^{6}$Li fusion reactions with closed channels}


\author{Wendi Chen$^{1}$, D.Y. Pang$^{1}$, Hairui Guo$^{2}$, Ye Tao$^{2}$, Weili Sun$^{2}$ and Yangjun Ying$^{2}$}
%\thanks{}
\affiliation{$^{1}$School of Physics, Beihang University, Beijing 100191, People’s Republic of China \\
$^{2}$ Institute of Applied Physics and Computational Mathematics, Beijing 100094, People’s Republic of China
}


\date{\today}

\begin{abstract}
Fusion reactions induced by the weakly bound nucleus $^{6}$Li with targets $^{28}$Si, $^{64}$Ni, $^{144}$Sm and $^{209}$Bi at energies around the Coulomb barrier are investigated within a three-body model where $^{6}$Li is described with an $\alpha + d$ cluster model. The total fusion (TF) cross sections are calculated with the continuum-discretized coupled-channel (CDCC) method and the complete fusion (CF) cross sections are extracted through the sum-rule model. The calculations demonstrate that (i) for the TF cross section calculations, the continuum states up to 40 MeV are found to be necessary, which corresponds to the inclusion of closed channels for light and medium mass targets, such as $^{28}$Si, $^{59}$Co and $^{144}$Sm, (ii) the converged CDCC results for TF cross section at energies above the Coulomb barrier are almost the same as single channel results in which the continuum coupling effect is neglected, and (iii) the continuum coupling strongly influences partial wave fusion cross sections and the closed channels play a significant role in the improvement of the description of the CF cross sections at energies below the Coulomb barrier for the $^6$Li+$^{28}$Si, $^{59}$Co and $^{144}$Sm systems.
\end{abstract}

%\keywords{}

\maketitle

\section{introduction}

Reactions induced by weakly bound projectiles have been extensively investigated in the last few decades\cite{Canto20061,Keeley2009396,Hagino2012,Canto20151,Lei2019}. In these reactions, breakup is a very important reaction channel and has a strong coupling effect on elastic scattering, inelastic scattering, transfer and fusion channels. In particular, special attention has been given to the continuum coupling effects on fusion reactions both theoretically and experimentally \cite{Beck2003,Dasgupta2004,Rath2009,Gautam2020121730,Chetna2022122418,Diaz-Torres2003,Camacho2019,Lubian2022}.
Among the stable weakly bound nuclei, $^{6}$Li has always been of interest, as it has only one bound state with a low separation energy of 1.47 MeV for breaking up into $\alpha$ and $d$ fragments. So far, many experiments for $^{6}$Li-induced fusion reactions have been reported, with targets varying from light nuclei, such as $^{27}$Al \cite{Barbara2007} and $^{28}$Si \cite{Sinha2017a}, to heavy nuclei, such as $^{197}$Au \cite{Palshetkar2004} and $^{209}$Bi \cite{Dasgupta2002,Dasgupta2004}. But the experimental data are still far from being fully understood.

Owing to the low binding energy of the projectile, there is a high probability that the projectile breaks into two or more fragments. In such cases, two kinds of fusion processes in the collisions of weakly bound nuclei have emerged from previous investigations, namely the complete fusion (CF) and the incomplete fusion (ICF). CF occurs when the whole weakly bound projectile is captured by the target. ICF occurs when some fragments of the projectile are captured and others escape. The sum of CF and ICF amounts to the total fusion (TF).

It is a great challenge to develop a realistic theory to describe the fusion process of weakly bound nuclei. Up to now, various theoretical approaches have been presented, among which the continuum discretized coupled-channel (CDCC) method has been the most popular one in making realistic predictions for the fusion reactions induced by weakly bound nuclei \cite{Hagino2000,Diaz-Torres2002,Diaz-Torres2003,Hashimoto2009,Parkar2016,Gomez2018,Camacho2019,Lei2019,Rangel2020,Lubian2022}, as it can effectively take the continuum coupling effect into account. For $^{6}$Li, many researchers \cite{Diaz-Torres2003,Gomez2018,Camacho2019} have adopted the CDCC method to calculate its TF cross sections. However, it is hard to evaluate the contributions from CF and ICF processes to TF cross sections for $^{6}$Li as its breakup fragments are both charged and their masses are comparable. Recently, a direct approach based on the CDCC method to evaluate the CF and ICF cross sections for $^{6,7}$Li fusion reactions was developed by Lubian et al. \cite{Rangel2020,Cortes2020,Lubian2022}, in which the probabilities of CF and ICF processes were obtained by integrating the scattering wave functions with different matrix elements of coupled-channel equations. They obtained a reasonable agreement between theory and experiment for $^{6,7}$Li fusion with heavy nuclei. On the other hand, Lei and Moro \cite{Lei2019} obtained the CF cross sections for $^{6,7}$Li+$^{209}$Bi indirectly by subtracting the cross sections of elastic breakup, nonelastic breakup and inelastic scattering from the total reaction cross section. Their results were in satisfactory agreement with experimental CF data, too. In addition, a different approach was developed by Parkar et al. \cite{Parkar2016} in which the CF, ICF and TF cross sections are obtained by three CDCC calculations with different short-range imaginary potentials. Their calculated results for $^{6,7}$Li+$^{198}$Pt and $^{209}$Bi were also in reasonable agreement with experimental data. These studies are all about 6Li fusion with heavy targets. Fusion reactions of 6Li with light and medium mass nuclei are still rare.

In many cases, the convergence of CDCC calculations is not easy to be achieved when they are applied in evaluating the fusion reaction cross sections. To overcome this problem, Diaz-Torres et al. \cite{Diaz-Torres2003} neglected the imaginary parts of off-diagonal couplings and only kept the imaginary parts in diagonal couplings in their study of $^{6}$Li fusion reactions. Similarly, Lubian et al. \cite{Lubian2022} neglected the imaginary parts of matrix elements between bound and breakup channels. In these works, the maximum energy of the continuum states, $\varepsilon_{max}$, for $^{6}$Li was set to be 6.0-8.0 MeV. It is far lower than the threshold energy of the continuum states, $\varepsilon_{th}$=$E_{cm}-\varepsilon_{b}$, where $E_{cm}$ is the incident energy in the centre of mass system and $\varepsilon_{b}$ is the separation energy of $\alpha$ and $d$ in the ground state $^6$Li.

It is well known that closed channels, whose channel energies are negative, play an important role in the low-energy reaction induced by weakly bound nuclei. Yahiro et al \cite{Yahiro1986} have discussed the coupling effect of closed channels on deuteron elastic scattering, which is visible for $d$+$^{58}$Ni reaction at low incident energy. Ogata and Yoshida \cite{Ogata2016} have reexamined the calculations for deuteron elastic breakup cross sections on $^{12}$C and $^{13}$Be at low incident energies. They pointed out that closed channels are required for CDCC calculations to obtain good agreement with the result of Faddeev-Alt-Grassberger-Sandhas theory, in which the three-body problem is exactly solved. Very recently, we \cite{Chen2022} presented a CDCC analysis for $^{6}$Li+$^{59}$Co reactions at energies around the Coulomb barrier and found that the $\varepsilon_{max}$ for $^{6}$Li should be at least 50.0 MeV to obtain converged elastic breakup reaction cross section. Therefore, it is worthy of reexamining the fusion cross section calculations by increasing $\varepsilon_{max}$ to check whether the continuum coupling effect is completely taken into account.

Furthermore, as a semi-classical approach, the sum-rule model \cite{Wilczynski1973386,Wilczynski1980,Wilczynski1982109,Mukeru2020121700,Mukeru2021} is adopted to distinguish the CF and ICF processes. According to this model, CF and ICF occur at lower and higher angular momenta respectively. It is of interest to study the continuum coupling effect on $^{6}$Li complete fusion with this model, as it can give a clear and simple dependence of CF cross sections on angular momenta and incident energy. Meanwhile, it can be applied to the $^6$Li fusion reactions with different targets, enabling us to investigate the influence of targets on CF.

In the present work, we study the coupling effect of continuum states on TF and CF cross sections for $^{6}$Li fusion with $^{28}$Si, $^{64}$Ni, $^{144}$Sm and $^{209}$Bi targets at energies around the Coulomb barrier. Particular attention is given to the reexamination of CDCC calculations for $^{6}$Li total fusion cross sections, where the numerical convergence problem should be solved by increasing $\varepsilon_{max}$. With the converged calculated results, the CF cross sections will be extracted by the sum-rule model to study the dependencies of CF process on breakup channels and targets. The coupling effects of open and closed channels will also be discussed.

The paper is organized as follows. The formalism is given in Sec. \ref{sec-2}. Reexamination of $^{6}$Li CDCC calculations for total fusion is presented in Sec. \ref{sec-3}. Sec. \ref{sec-4} discusses the coupling effect of continuum states on TF and CF cross sections in detail. Finally, the summary and conclusion are given in Sec. \ref{sec-5}.

\section{Theoretical model} \label{sec-2}

\subsection{$\alpha$+$d$ cluster model for $^6$Li}

$^{6}$Li is described in $\alpha$+$d$ cluster model. Its internal wave function is written as
\begin{equation}\label{e-wf}
  \psi =\mathcal{A} \left\{ \varphi \left( \alpha \right) \left[ \varphi _s\left( d \right) \otimes \chi _l\left( \vec{r} \right) \right] _{I}^{M} \right\} ,
\end{equation}
where $\mathcal{A}$ means the antisymmetrization of the nucleons. $\varphi (\alpha)$ and $\varphi _s (d)$ are the intrinsic wave functions of $\alpha$ and $d$ clusters respectively. $\vec{r}$ is the relative coordinate between two clusters. $\chi _l$ represents the $\alpha-d$ relative motive with angular momentum $l$, which couples with the spin of deuteron cluster, $s$, to form the total spin $I$ and its projection $M$.

\begin{table}[tbp]
\caption{Parameters of the effective $\alpha-d$ potential $V_{\alpha-d}$ \cite{Chen2022} in Woods-Saxon form. This potential is $l$-dependent. Its nuclear radius $R_N$ and Coulomb radius $R_C$ are both 2.1 fm. The diffuseness parameter $a$ is 0.65 fm. $V_0$ and $V_{so}$ represent the depths of the central and spin-orbit potentials respectively and their units are MeV.}
\label{table-pot}
\begin{ruledtabular}
\begin{tabular}{cccc}
 $l$ & 0 & 1 & 2 \\
\hline
 $V_0$    & 67.69 & 63.90 & 65.33 \\
 $V_{so}$ &  0.00 &  5.72 &  4.78 \\
\end{tabular}
\end{ruledtabular}
\end{table}

\begin{table}[tbp]
\caption{Calculated resonance energies $\varepsilon_{res}^{cal}$ and widths $\Gamma_{res}^{cal}$ compared with experimental value $\varepsilon_{res}^{exp}$ and $\Gamma_{res}^{exp}$ \cite{Tilley20023}. Their units are MeV.}
\label{table-res}
\begin{ruledtabular}
\begin{tabular}{cccccc}
 State & $\varepsilon_{res}^{cal}$ & $\Gamma_{res}^{cal}$ & $\varepsilon_{res}^{exp}$ & $\Gamma_{res}^{exp}$ \\
\hline
 3$^+$ & 0.710 & 0.084 & 0.716 & 0.024 \\
 2$^+$ & 3.00  & 1.12  & 2.84  & 1.30  \\
 1$^+$ & 4.24  & 2.93  & 4.18  & 1.50  \\
\end{tabular}
\end{ruledtabular}
\end{table}

\begin{figure}[htbp]
  \centering
  \includegraphics[width=\columnwidth]{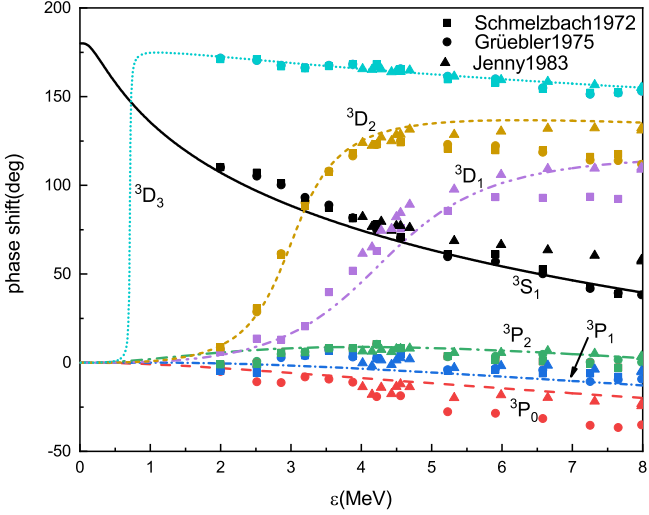}
  \caption{(Color online) Calculated $\alpha-d$ scattering phase shift with $V_{\alpha-d}$ \cite{Chen2022}, shown as a function of the relative $\alpha-d$ energy in centre of mass system, $\varepsilon$. Experimental data are taken from Schmelzbach et al. \cite{Schmelzbach1972} (squares), Gr$\mathrm{\ddot{u}}$ebler et al. \cite{Gruebler1975} (circles) and Jenny et al. \cite{Jenny1983} (triangles).}
  \label{fig-ps}
\end{figure}

In the present work, a simple version of the orthogonality condition model \cite{Saito1969,Horiuchi1977,Sakuragi1986,Guo2014} is used to calculate the $^6$Li internal wave function $\psi$. The effects of the antisymmetrization of nucleons are taken into account approximately by employing an effective $\alpha-d$ potential, $V_{\alpha-d}$, and excluding the deepest bound state as the forbidden state. Therefore, $\psi$ is calculated by solving a Schrodinger equation with $V_{\alpha-d}$ \cite{Chen2022}, which is $l$-dependent. $V_{\alpha-d}$ are parameterized by the Woods-Saxon form \cite{Koning2003231}, including the central and spin–orbit potentials. Its parameters for $l$=0, 1 and 2 are listed in Table \ref{table-pot}. $V_{\alpha-d}$ can well reproduce the binding energy of 1.47 MeV ($l$=0, $s$=1 and $I^{\pi}$=$1^{+}$) as well as the $3^{+}$, $2^{+}$ and $1^{+}$ resonance states in $D$-wave continuum. The calculated resonance energies and widths are shown in Table \ref{table-res}, compared with the experimental data \cite{Tilley20023}. $V_{\alpha-d}$ also describes the low-energy $\alpha-d$ scattering phase shifts well, shown in Fig \ref{fig-ps} as a function of the relative $\alpha-d$ energy in centre of mass system $\varepsilon$.

\begin{figure}[htbp]
  \centering
  \includegraphics[width=\columnwidth]{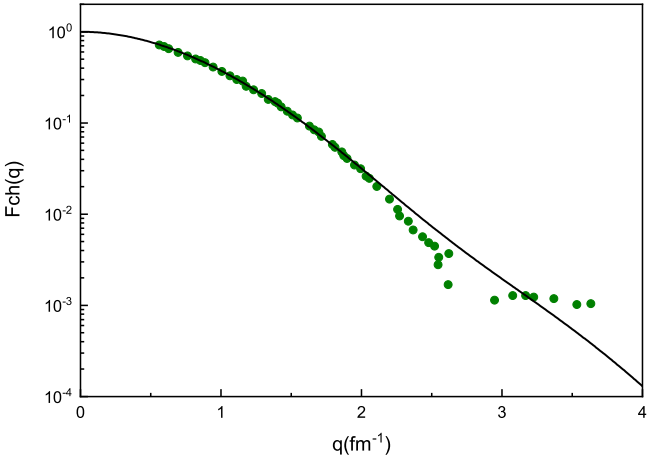}
  \caption{(Color online) Calculated charge form factors of elastic electron scattering by $^6$Li, compared with the experimental data \cite{Bergstrom1979458}.}
  \label{fig-Fch}
\end{figure}

The spacial wave functions of the $d$ and $\alpha$ clusters are assumed to be $(1s)^2$ and $(1s)^4$ harmonic oscillator shell model wave functions with different oscillator constants $\beta_{d}$ and $\beta_{\alpha}$ respectively ($\beta =m\omega /\hbar$), and expressed as
\begin{equation}\label{e-wf-da}
\begin{aligned}
\varphi \left( d \right) &=N_d\exp \left[ -\frac{\beta _d}{2}\sum_{i\in d}{\left( \vec{r}_i-\vec{R}_d \right) ^2} \right] , \\
\varphi \left( \alpha \right) &=N_{\alpha}\exp \left[ -\frac{\beta _{\alpha}}{2}\sum_{i\in \alpha}{\left( \vec{r}_i-\vec{R}_{\alpha} \right) ^2} \right]
\end{aligned}
\end{equation}
where $\beta_d$=0.390 fm$^{-2}$ and $\beta_{\alpha}$=0.375 fm$^{-2}$. $\vec{r_i}$ is the coordinate of particle $i$ in $^6$Li relative to the centre of mass of $^6$Li. $\vec{R}_d$ and $\vec{R}_{\alpha}$ represent the centre of mass for the corresponding clusters. $N_{d}$ and $N_{\alpha}$ are the corresponding normalized factors. The calculated value of root-mean-square matter radius and that of charge radius are both 2.54 fm, which agree with the experimental value 2.54$\pm$0.03 fm \cite{Tanihata1985} (the measured values for these two radii are the same). The charge form factor of elastic electron scattering by $^6$Li is also calculated, as shown in Fig. \ref{fig-Fch}. Reasonable agreement is obtained with the experimental data \cite{Bergstrom1979458}.

\subsection{The CDCC method and the sum-rule model}

Detailed CDCC formalism can be found in Refs. \cite{Sakuragi1986,Austern1987125,Diaz-Torres2002,Chen2022}. In this method, a finite number of discretized and square-integrable states are adopted to represent the continuum states of the projectile. Hence, the total wave function with total angular momentum $J$ and parity $\pi$ can be expressed as
\begin{equation}\label{e-totalwf}
  \varPsi ^{J\pi} =\sum_{\beta}{\varPhi _{\beta} ^{J\pi} \psi _{\beta}}
\end{equation}
where $\beta$ represents the reaction channel. $\psi _{\beta}$ is the internal wave function of $^6$Li in $\beta$ channel. The wave functions for bound and discretized states are expressed on the same footing. $\varPhi _{\beta} ^{J\pi}$ is the relative motion wave function between the $^6$Li in $\beta$ channel and target. $\varPhi _{\beta} ^{J\pi}$ is determined by the coupled-channel equations
\begin{equation}\label{e-CC}
  \left[ T+E_{\beta}-U_{\beta \beta} ^{J\pi} \right] \varPhi _{\beta} ^{J\pi}=-\sum_{\beta ^ \prime \ne \beta}{U_{\beta \beta ^ \prime} ^{J\pi} }\varPhi _{\beta ^ \prime} ^{J\pi},
\end{equation}
where $E_{\beta}$ is the channel energy. $T$ denotes the kinetic energy of $^6$Li-target relative motion. The coupling matrix element $U_{\beta \beta ^ \prime} ^{J\pi}$ is calculated as
\begin{equation}\label{e-Vcc}
U_{\beta \beta ^ \prime} ^{J\pi}=\left< \psi _{\beta} \right|U_{\alpha -T}+U_{d-T}\left| \psi _{\beta ^ \prime} \right>,
\end{equation}
where $U_{\alpha -T}$ and $U_{d-T}$ are the optical potentials for $\alpha$ and $d$ with the target respectively. The imaginary parts of the optical potentials, $W_{\alpha -T}$ and $W_{d -T}$, are responsible for the absorption of $^6$Li by the target. Therefore, the coupling matrix elements corresponding to the absorption is given by
\begin{equation}\label{e-Wcc}
  W_{\beta \beta ^ \prime}^{J\pi}=\left< \psi _{\beta} \right|W_{\alpha -T}+W_{d-T}\left| \psi _{\beta ^ \prime} \right>.
\end{equation}
The total fusion cross section is then obtained as
\begin{equation}\label{e-TF}
  \sigma _{TF}=\sum_{J=0}^{J_{\max}}{\sigma _{J}},
\end{equation}
where
\begin{equation}\label{e-TFJ}
  \sigma _J=\frac{K}{E}\sum_{\pi \beta \beta ^{\prime}}{\left< \varPhi _{\beta}^{J\pi} \right|W_{\beta \beta ^{\prime}}^{J\pi}\left| \varPhi _{\beta ^{\prime}}^{J\pi} \right>}.
\end{equation}
$K$ is the projectile-target relative wave number in the incident channel. It should be emphasized that the wave functions of all channels, including open and closed channels, are required in the calculations for $\sigma _J$ with Eq. (\ref{e-TFJ}). This method is equivalent to the computing approach with $S$ matrix \cite{Thompson1988167}, in which only the $S$ matrices of open channels are used.

According to the sum-rule model, the complete fusion cross section can be extracted directly as
\begin{equation}\label{e-CF}
  \sigma _{CF}=\sum_{J=0}^{J_{c}}{\sigma _{J}},
\end{equation}
where $J_c$ is the cut-off angular momentum (see Sec. \ref{sec-4-2} for details).

\section{Convergence of total fusion cross section calculations} \label{sec-3}

Optical potential with a short-range imaginary part can be applied to calculate the fusion cross section, which is equivalent to the use of an incoming boundary condition inside the Coulomb barrier \cite{Rhoadesbrown198419}. In principle, the calculated fusion cross section should be independent of the imaginary part of the optical potential.

In this work, the S$\tilde{a}$o Paulo potential version 2 (SPP2) \cite{Chamon2021108061} is adopted for the real parts of the nuclear interactions between the $\alpha$ and $d$ clusters with the target. SPP2 potential is calculated with the double folding method, which requires the nucleon distributions of projectile and target. We adopt the theoretical nuclear distribution embedded in the code for the target, which is calculated by an axially-symmetric self-consistent Dirac-Hartree-Bogoliubov mean field approach \cite{Carlson2000}. The nucleon distributions for $\alpha$ and $d$ clusters are calculated with the wave function in Eq. (\ref{e-wf-da}). For the Coulomb potentials, the radius factor $r_C$ for two clusters with the target are both set to be 1.5 fm.

The short-range imaginary parts of the $d$-target and $\alpha$-target optical potentials are parameterized in the Woods-Saxon form. Their radius factor $r_W$ and diffuseness parameter $a_W$ are set to be 0.8 fm and 0.1 fm respectively so that the imaginary parts are inside the Coulomb barrier completely. The depths of the two imaginary parts $W_0$ are set to be the same, varying from 20.0 MeV to 80.0 MeV to examine the convergence of total fusion cross section calculation.

\begin{figure}[htbp]
  \centering
  \includegraphics[width=\columnwidth]{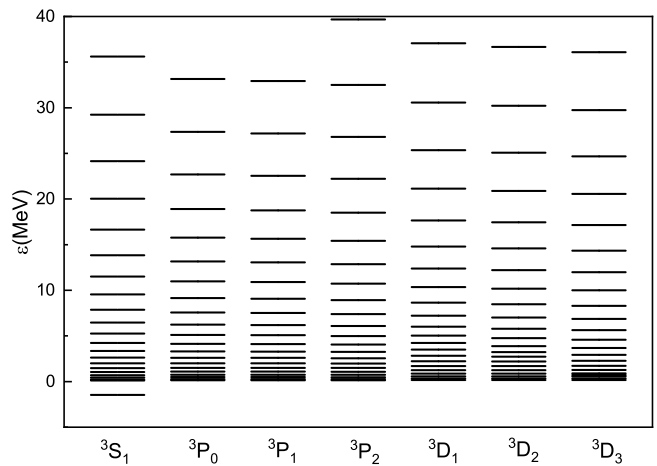}
  \caption{Energy spectrum of $^6$Li bound and pseudo-states up to $D$-wave continuum and 40.0 MeV, calculated with regularized Lagrange–Laguerre mesh method ($N$=35 and $h$=0.5 fm).}
  \label{fig-e-sample}
\end{figure}

Following our previous study \cite{Chen2022}, the internal Hamiltonian of $^6$Li is diagonalized by the regularized Lagrange–Laguerre mesh method \cite{Druet201088,Druet2012,Baye20151}. This way of discretizing the continuum states is called the pseudo-state method, which diagonalizes the Hamiltonian by square integrable basis functions and can generate the bound and pseudo-states together. The pseudo-states are used to represent the continuum states. The basis functions are defined as
\begin{equation}\label{e-basis}
f_i\left( r \right) =\frac{\left( -1 \right) ^i}{\sqrt{hx_i}}\frac{L_N\left( r/h \right)}{r-hx_i}re^{-r/2h},i=1,2,...,N,
\end{equation}
where $L_N$ is the Laguerre polynomial of degree $N$. $x_i$ corresponds to the zeros of $L_N$, that is
\begin{equation}\label{e-zeros}
  L_N(x_i)=0, i=1,2,...,N.
\end{equation}
$h$ is a scaling parameter, adopted to the typical size of the system. For more details one can see Refs. \cite{Druet201088,Druet2012,Baye20151,Chen2022}.

In this paper, the number of basis function $N$ and the scaling parameter $h$ are set to be 35 and 0.5 fm respectively. Many tests have been performed to ensure that the calculated fusion cross sections are insensitive to the parameters $N$ and $h$. Fig. \ref{fig-e-sample} shows the energy spectrum of $^6$Li bound and pseudo-states up to $D$-wave continuum and 40.0 MeV.

\begin{figure}[htbp]
  \centering
  \includegraphics[width=\columnwidth]{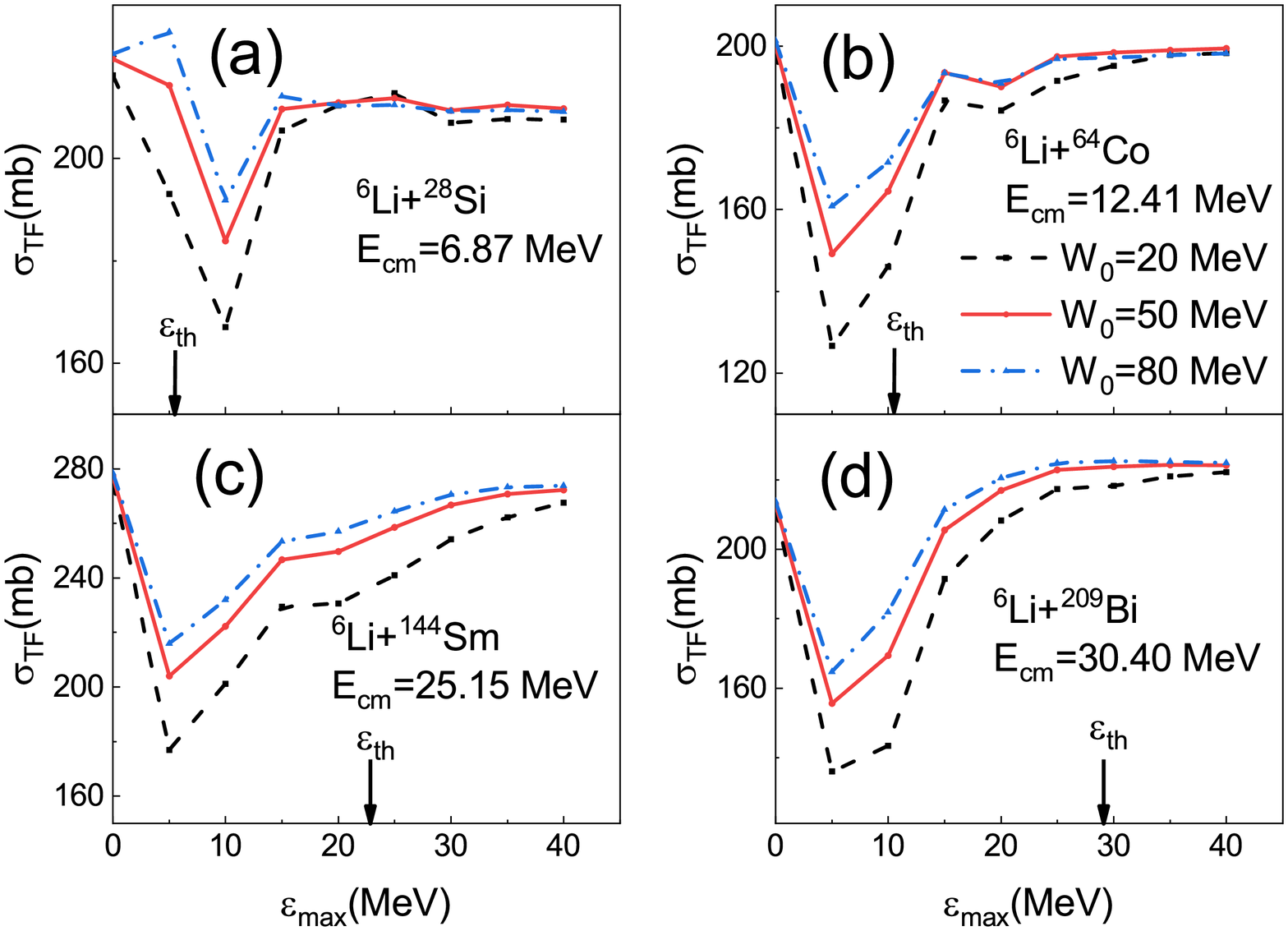}
  \caption{(Color online) Calculated TF cross sections for $^6$Li+$^{28}$Si, $^{64}$Ni, $^{144}$Sm and $^{209}$Bi systems at $E_{cm}=V_B$ with different imaginary part depth $W_0$ and maximum energy of the continuum state $\varepsilon_{max}$. $V_B$ represents the Coulomb barrier. The arrow indicates the threshold energy of the continuum states $\varepsilon_{th}$=$E_{c.m.}-1.47$ MeV. See text for details.}
  \label{fig-TF-VB-convergence}
\end{figure}

We firstly perform calculations for $^6$Li+$^{28}$Si, $^{64}$Ni, $^{144}$Sm and $^{209}$Bi systems at incident energy in center of mass system $E_{cm}=V_B$, where $V_B$ is the Coulomb barrier measured in the references. $V_B$=6.87, 12.41, 25.15 and 30.40 MeV for $^6$Li+$^{28}$Si \cite{Sinha2017a}, $^{64}$Ni \cite{Shaikh2014}, $^{144}$Sm \cite{Rath2009} and $^{209}$Bi \cite{Dasgupta2002} systems respectively. $l$=0, 1 and 2 states are taken into CDCC calculations. The calculations are made with the maximum continuum energy $\varepsilon_{max}$ from 0 to 40 MeV for all partial waves with the depth of the imaginary potential being 20, 50 and 80 MeV, respectively. The results are shown in Fig. 4. Form these results, we can get the following important information:

\begin{enumerate}

\item The maximum continuum energy set in the calculations $\varepsilon_{max}$ should be large enough to ensure that the $\sigma_{TF}$ do not depend on the choice of the imaginary potential depth. On the other hand, when $\varepsilon _{max}$ is set large enough, the resulting $\sigma_{TF}$ values do converge to a value which is independent of the depths of the imaginary potentials;

\item For $^6$Li+$^{28}$Si, $^{64}$Ni and $^{144}$Sm systems, the inclusion of closed channels is necessary for the convergence of the total fusion cross sections.

\item For $^6$Li+$^{209}$Bi system, the $\sigma_{TF}$ calculated with $\varepsilon_{max}$=$\varepsilon_{th}$ only differs from that calculated with $\varepsilon_{max}$=40.0 MeV by 0.3$\%$ when $W_0$=50 MeV. The coupling effect of closed channels is weak for $^6$Li total fusion with such a heavy nucleus.

\end{enumerate}

Practically, $\varepsilon_{max}=40$ MeV seems to be sufficient for all the four reaction systems. In this case, the changes in the total fusion cross sections are less than 2\% when the imaginary potential depth changes from 20 to 80 MeV. This $\varepsilon_{max}$ is well above the threshold energy of the continuum state $\varepsilon_{th}$=$E_{c.m.}-1.47$ MeV and the closed channels are well included in calculations.

For a further understanding of how the cut-off of continuum state energy and the imaginary part of optical potentials influence the fusion cross section calculations, we compare the radial relative wave function for $^6$Li+$^{64}$Ni system at $E_{c.m.}=V_B$ with different $\varepsilon_{max}$ and $W_0$. The radial relative wave function is defined as
\begin{equation}\label{e-UR}
  u^J_{\beta}(R)=R\Psi ^J_{\beta}(R),
\end{equation}
where $R$ is the relative distance between the centers of mass of $^6$Li and target. To explicate the wave functions, the orbital angular momentums of the projectile-target relative motive in incoming and outgoing channels are required, namely $L$ and $L'$. The radial wave functions at $J^{\pi}$=$1^+$, $4^-$ and $10^+$ are calculated for comparision. The partial wave fusion cross section $\sigma_J$ reaches a maximum at $J$=4, which is also the cut-off angular momentum for this reaction (See Sec. \ref{sec-4-2} for more details).

\begin{figure}[htbp]
  \centering
  \includegraphics[width=\columnwidth]{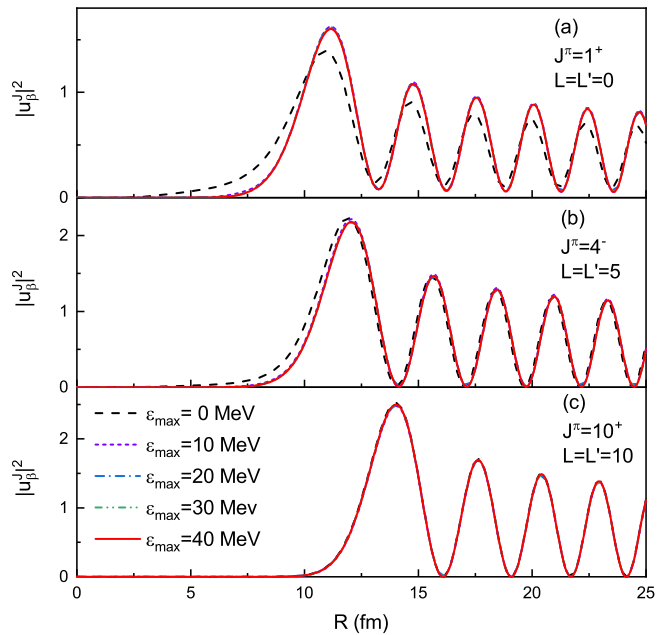}
  \caption{(Color online) Square of the absolute value of the radial wave functions, $| u_{\beta}^{J} |^2$, of elastic scattering channel for $^6$Li+$^{64}$Ni reaction at $E_{c.m.}=V_B=$ 12.41 MeV. The wave functions calculated with $\varepsilon_{max}$=0, 10, 20, 30 and 40 MeV are represented by dashed, short dashed, dash-dotted, dash-double-dotted and solid lines respectively. The subfigures (a), (b) and (c) show the $\left| u_{\beta}^{J} \right|^2$ at $J^{\pi}$=$1^+$, $4^-$ and $10^+$ with $L$=$L'$=0, 5 and 10 respectively. See text for details.}
  \label{fig-6+64-wf-elastic}
\end{figure}

In the present work, we study the influence of $\varepsilon_{max}$ and $W_0$ on wave functions of elastic scattering, $3^+$ resonance breakup and closed channels. The $3^+$ resonance state of $^6$Li is represented by the pseudo-state in $^{3}D_3$ partial wave with eigen energy 0.7111 MeV. For the closed channel wave functions, the $u^J_{\beta}$ of the pseudo-state in $^{3}S_1$ partial wave with eigen energy 11.5108 MeV are calculated.

As the wave functions are used to calculate fusion cross sections as shown in Eq. (\ref{e-TFJ}), we focus on the square of the absolute value of wave functions $| u_{\beta}^{J} |^2$. With a fixed $W_0$=50 MeV, the wave functions are calculated with $\varepsilon_{max}$=0, 10, 20, 30 and 40 MeV. Fig. \ref{fig-6+64-wf-elastic} shows $| u_{\beta}^{J} |^2$ of elastic scattering. $L$=$L'$=0, 5 and 10 for $J^{\pi}$=$1^+$, $4^-$ and $10^+$ respectively. It is observed that $| u_{\beta}^{J} |^2$ well converges with $\varepsilon_{max}$=10 MeV, which coincides with the general adopted $\varepsilon_{max}$ for obtaining converged $^6$Li elastic scattering angular distributions.

\begin{figure}[htbp]
  \centering
  \includegraphics[width=\columnwidth]{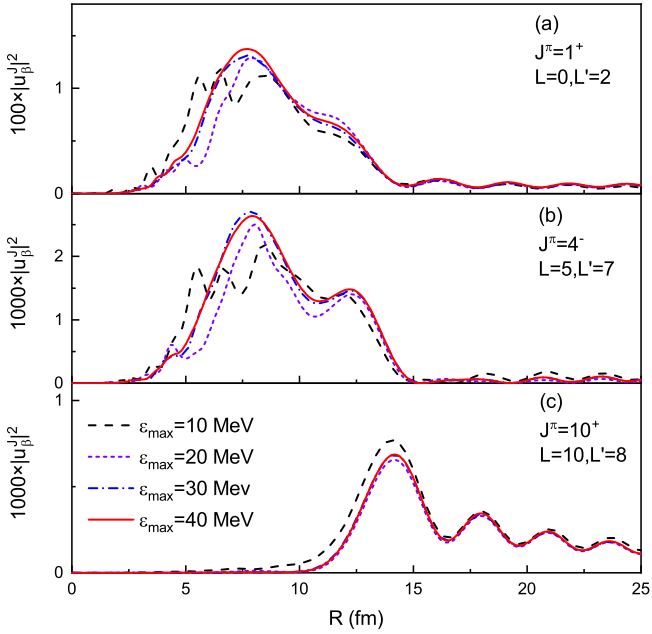}
  \caption{(Color online) Same as Fig. \ref{fig-6+64-wf-elastic} but for $3^+$ resonance breakup. The wave functions calculated with $\varepsilon_{max}$=10, 20, 30 and 40 MeV are denoted by dashed, short dashed, dash-dotted and solid lines respectively. $L$=0, 5 and 10, $L'$=2, 7 and 8 for $J^{\pi}$=$1^+$, $4^-$ and $10^+$ respectively. The $| u_{\beta}^{J} |^2$ in subfigures (a), (b) and (c) are multiplied by 100, 1000 and 1000 respectively.}
  \label{fig-6+64-wf-res}
\end{figure}

The $| u_{\beta}^{J} |^2$ for $3^+$ resonance breakup are plotted in Fig. \ref{fig-6+64-wf-res} for comparison. $L$=0, 5 and 10, $L'$=2, 7 and 8 for $J^{\pi}$=$1^+$, $4^-$ and $10^+$ respectively. Different from the elastic scattering, the convergence of $| u_{\beta}^{J} |^2$ for $3^+$ resonance breakup needs a pretty high $\varepsilon_{max}$=40 MeV, especially in the inner region ($R <$ 15 fm).

\begin{figure}[htbp]
  \centering
  \includegraphics[width=\columnwidth]{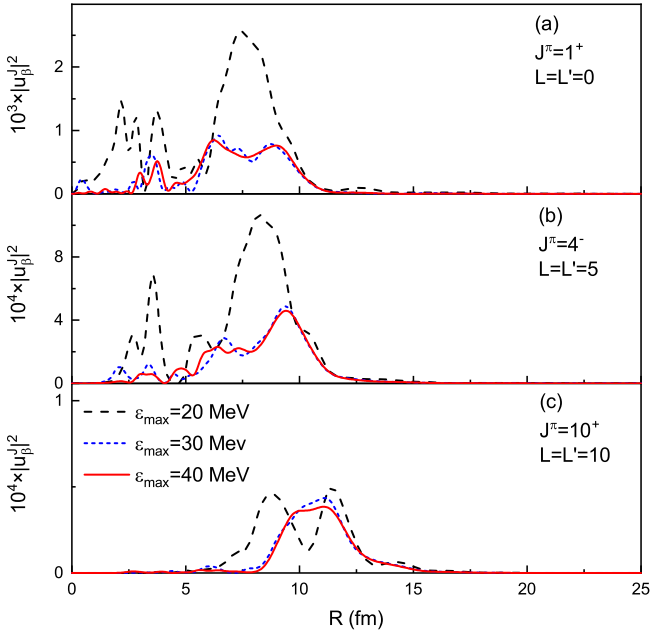}
  \caption{(Color online) Same as Fig. \ref{fig-6+64-wf-elastic} but for the closed channel. The wave functions calculated with $\varepsilon_{max}$=20, 30 and 40 MeV are denoted by dashed, short dashed and solid lines respectively. $L$=$L'$=0, 5 and 10 for $J^{\pi}$=$1^+$, $4^-$ and $10^+$ respectively. The $| u_{\beta}^{J} |^2$ in subfigures (a), (b) and (c) are multiplied by 10$^3$, 10$^4$ and 10$^4$ respectively.}
  \label{fig-6+64-wf-close}
\end{figure}

In Fig. \ref{fig-6+64-wf-close}, the $| u_{\beta}^{J} |^2$ of the closed channels are shown. Although the wave functions are not completely converged with $\varepsilon_{max}$=40 MeV, we found that these small differences between the closed channel wave functions calculated with $\varepsilon_{max}$=30 and 40 MeV have little effect on the fusion cross sections.

\begin{figure}[tbp]
  \centering
  \includegraphics[width=\columnwidth]{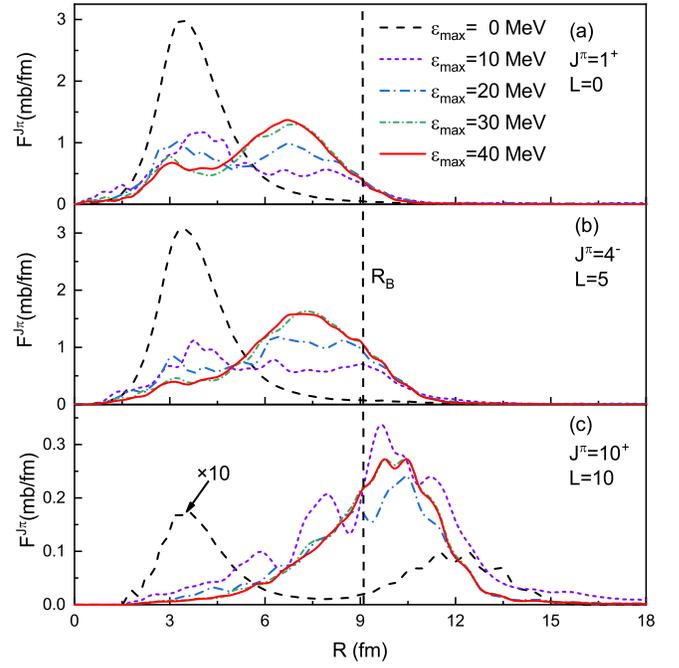}
  \caption{(Color online) The integrands $F$ for $^6$Li+$^{64}$Ni reaction at $E_{c.m.}=V_B=$ 12.41 MeV. The integrands calculated with $\varepsilon_{max}$=0, 10, 20, 30 and 40 MeV are denoted by dashed, short dashed, dash-dotted, short dash-dotted and solid lines respectively. They are evaluated at (a) $J^{\pi}$=$1^+$, $L$=0, (b) $J^{\pi}$=$4^-$, $L$=5 and (a) $J^{\pi}$=$10^+$, $L$=10. $R_B$=9.1 fm is the Coulomb barrier radius \cite{Shaikh2014}. The $F$ calculated with $\varepsilon_{max}$=0 at $J^{\pi}$=$10^+$ is multiplied by 10 for easier viewing.}
  \label{fig-6+64-F-emax}
\end{figure}

In Eq. (\ref{e-TFJ}), the partial wave fusion cross section are calculated in an integration method. We define
\begin{equation}\label{e-F-function}
  F^{J\pi}(R)=\frac{K}{E}\sum_{\beta \beta ^{\prime}}{\varPhi _{\beta}^{J\pi*}\left( R \right)}W_{\beta \beta \prime}^{J\pi}\left( R \right) \varPhi _{\beta \prime}^{J\pi}\left( R \right).
\end{equation}
 Therefore
\begin{equation}\label{e-TFJ-F-inte}
  \sigma _J=\sum_{\pi}{\int{F^{J\pi}(R) dR}}.
\end{equation}
One point worth emphasizing is that $F^{J\pi}$ is dependent on $J$, $\pi$ and $L$. The summation of $L$ is necessary for the projectile with non-zero spin in the calculation of $\sigma _J$, but we omit it in Eq. (\ref{e-TFJ-F-inte}) for simplification. As the range of $W_{\beta \beta ^ \prime}^{J\pi}$ is no more than 15 fm in general, it can be deduced that $F$ is sensitive to the wave functions in the inner part and the convergence of $F$ also requires a high $\varepsilon_{max}$.

The $F^{J\pi}$ at $J^{\pi}$=$1^+$, $4^-$ and $10^+$ are plotted in Fig. \ref{fig-6+64-F-emax} with $L$=0, 5 and 10 respectively. Convergence within the deviation of less than 2$\%$ is reached with $\varepsilon_{max}$=40 MeV. The $F^{J\pi}$ calculated with $\varepsilon_{max}$=0 and 40 MeV have the same order of magnitude at $J^{\pi}$=$1^+$ and $4^-$, but the former becomes about 10 times less than the latter at $J^{\pi}$=$10^+$. We multiply the $F^{J\pi}$ calculated with $\varepsilon_{max}$=0 at $J^{\pi}$=$10^+$ by 10 for easier viewing. A more detailed discussion about the influence of $\varepsilon_{max}$ on $\sigma_J$ will be given in Sec. \ref{sec-4-2}. On the other hand, the maximum of $F^{J\pi}$ locates at $R$=3.4 fm when $\varepsilon_{max}$=0 and it moves outwards with the increase of $\varepsilon_{max}$. The peak of converged $F^{J\pi}$ at $J^{\pi}$=$10^+$ even locates outside of the Coulomb barrier radius $R_B$=9.1 fm \cite{Shaikh2014}. The above comparisons indicate that the continuum coupling effect not only increases the partial wave fusion cross sections at relatively higher partial waves but also fully changes the fusion mechanism.

\begin{figure}[tbp]
  \centering
  \includegraphics[width=\columnwidth]{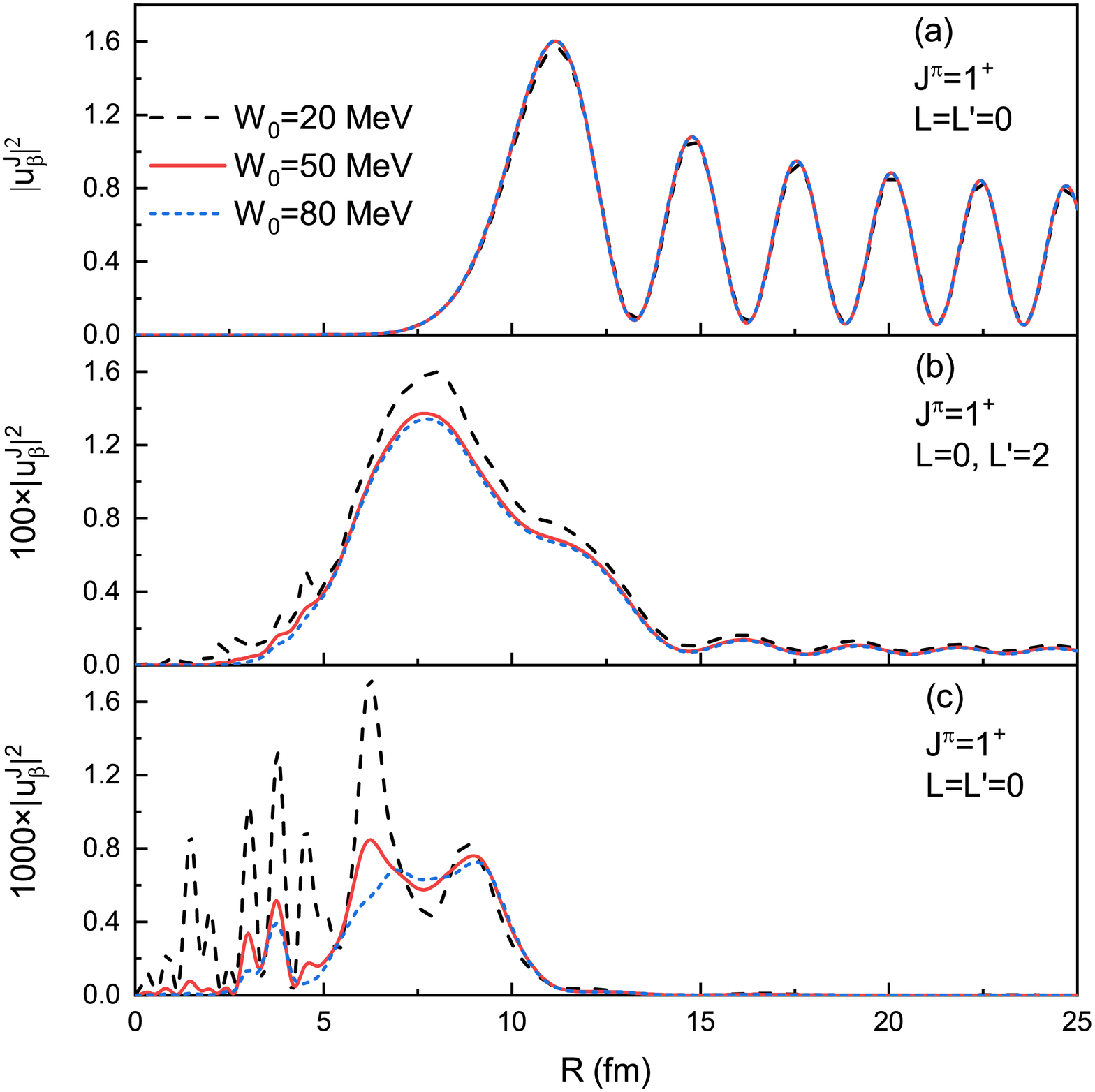}
  \caption{(Color online) Square of the absolute value of the radial wave functions, $| u_{\beta}^{J} |^2$, for $^6$Li+$^{64}$Ni reaction at $E_{c.m.}=V_B=$ 12.41 MeV and $J^{\pi}$=$1^+$. The wave functions calculated with $W_0$=20, 50 and 80 MeV are represented by dashed, solid, and short dashed lines respectively. The subfigures (a), (b) and (c) show the $| u_{\beta}^{J} |^2$ of elastic scattering, $3^+$ resonance breakup and closed channel respectively. The $| u_{\beta}^{J} |^2$ of $3^+$ resonance breakup and closed channel are multiplied by 100 and 1000 respectively. See text for details.}
  \label{fig-6+64-wf-1+}
\end{figure}

Another worth-considering issue is that the total fusion cross section is independent of the imaginary part of optical potentials when $\varepsilon_{max}$ is large enough. With a fixed $\varepsilon_{max}$=40 MeV, we investigate how the $u^J_{\beta}$ changes as the potential depth of the imaginary part varies. Fig. \ref{fig-6+64-wf-1+} shows the $| u_{\beta}^{J} |^2$ calculated with $W_0$=20, 50 and 80 MeV at $J^{\pi}$=$1^+$. As $W_0$ increases from 20 to 80 MeV, the $| u_{\beta}^{J} |^2$ of elastic scattering have hardly any changes, while the $| u_{\beta}^{J} |^2$ of $3^+$ resonance breakup and closed channel vary visibly and their values are negatively correlated with $W_0$ in the very inner regions $R <$ 4 and 7 fm respectively. For example, the $| u_{\beta}^{J} |^2$ of closed channel calculated with $W_0$=20, 50 and 80 MeV all reach a peak at $R$=3.75 fm and their values are 0.0013, 0.0005 and 0.0003 respectively.

\begin{figure}[tbp]
  \centering
  \includegraphics[width=\columnwidth]{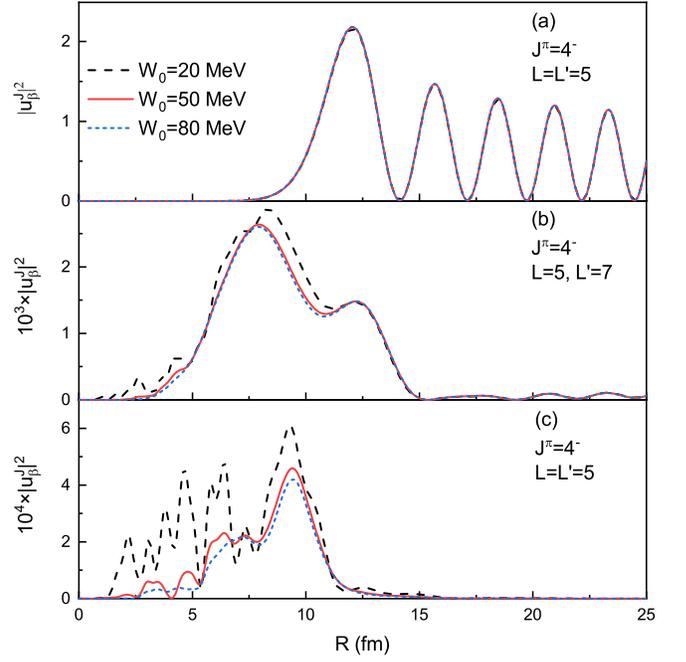}
  \caption{(Color online) Same as Fig. \ref{fig-6+64-wf-1+} but for $J^{\pi}$=$4^-$. The $| u_{\beta}^{J} |^2$ of $3^+$ resonance breakup and closed channel are multiplied by 10$^3$ and 10$^4$ respectively.}
  \label{fig-6+64-wf-4-}
\end{figure}

\begin{figure}[tbp]
  \centering
  \includegraphics[width=\columnwidth]{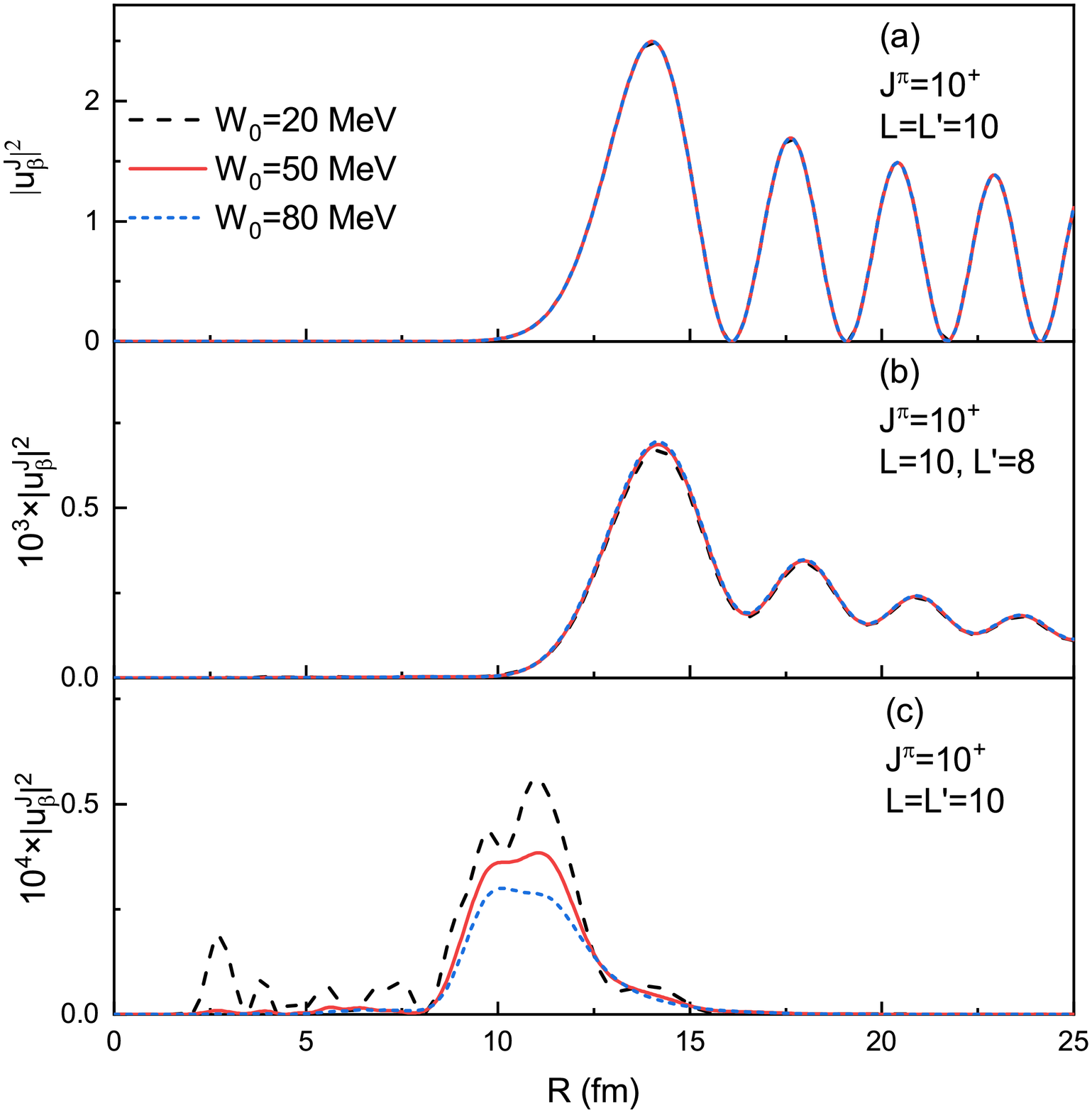}
  \caption{(Color online) Same as Fig. \ref{fig-6+64-wf-1+} but for $J^{\pi}$=$10+$. The $| u_{\beta}^{J} |^2$ of $3^+$ resonance breakup and closed channel are multiplied by 10$^3$ and 10$^4$ respectively.}
  \label{fig-6+64-wf-10+}
\end{figure}

Comparisons at $J^{\pi}$=$4^-$ and $10^+$ are presented in Figs. \ref{fig-6+64-wf-4-} and \ref{fig-6+64-wf-10+}. The $| u_{\beta}^{J} |^2$ of elastic scattering is stable against $W_0$. The value of $| u_{\beta}^{J} |^2$ of $3^+$ resonance breakup is still negatively correlated with $W_0$ at $J^{\pi}$=$4^-$ in the region $R <$ 4 fm, but it becomes independent of $W_0$ at $J^{\pi}$=$10^+$. However, for the $| u_{\beta}^{J} |^2$ of closed channel, the negative correlation with $W_0$ is kept at both $J^{\pi}$=$4^-$ and $10^+$.

\begin{figure}[tbp]
  \centering
  \includegraphics[width=\columnwidth]{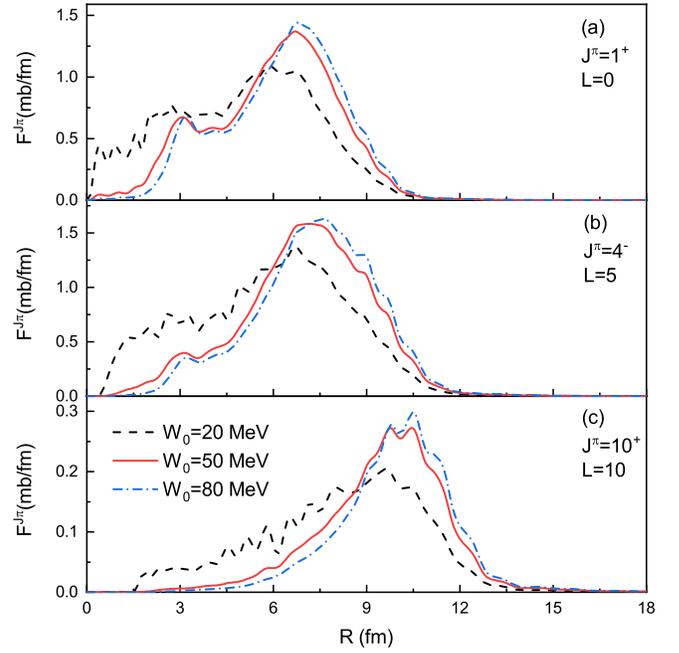}
  \caption{(Color online) The integrands $F$ for $^6$Li+$^{64}$Ni reaction at $E_{c.m.}=V_B=$ 12.41 MeV. The integrands calculated with $W_0$=20, 50 and 80 MeV are denoted by dashed, solid and dash-dotted lines respectively. See text for details.}
  \label{fig-6+64-F-W0}
\end{figure}

\begin{figure}[tbp]
  \centering
  \includegraphics[width=\columnwidth]{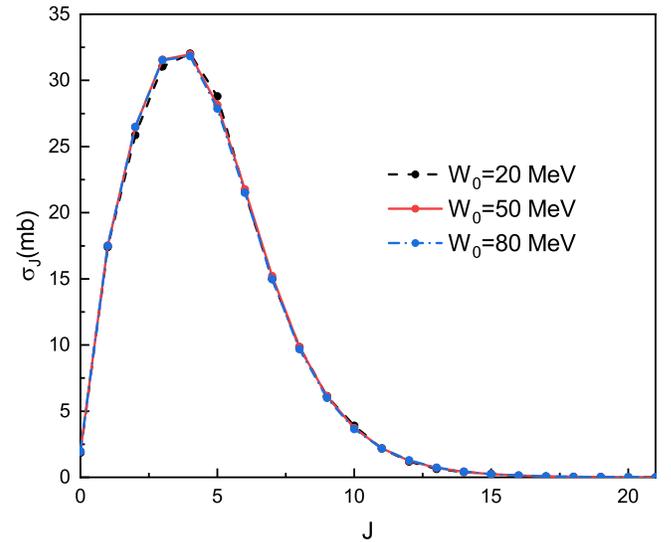}
  \caption{(Color online) The partial wave fusion cross section $\sigma_J$ for $^6$Li+$^{64}$Ni reaction at $E_{c.m.}=V_B=$ 12.41 MeV. The $\sigma_J$ with $W_0$=20, 50 and 80 MeV are denoted by dashed, solid and dash-dotted lines respectively. See text for details.}
  \label{fig-6+64-CF-J-W0}
\end{figure}

Fig. \ref{fig-6+64-F-W0} shows the integrands $F^{J\pi}$ calculated with $W_0$=20, 50 and 80 MeV at $J^{\pi}$=$1^+$, $4^-$ and $10^+$. The $F^{J\pi}$ calculated with $W_0$=20 MeV is flatter than those calculated with $W_0$=50 and 80 MeV. Fortunately, although the three kinds of $F$ do not coincide perfectly, their integrations are almost the same, which will not affect the subsequent analysis. Furthermore, an overall comparison of partial wave fusion cross sections calculated with different $W_0$ is presented in Fig. \ref{fig-6+64-CF-J-W0}. $\sigma_J$ at each $J$ is indeed independent of $W_0$ when $\varepsilon_{max}$=40 MeV.

\begin{figure}[tbp]
  \centering
  \includegraphics[width=\columnwidth]{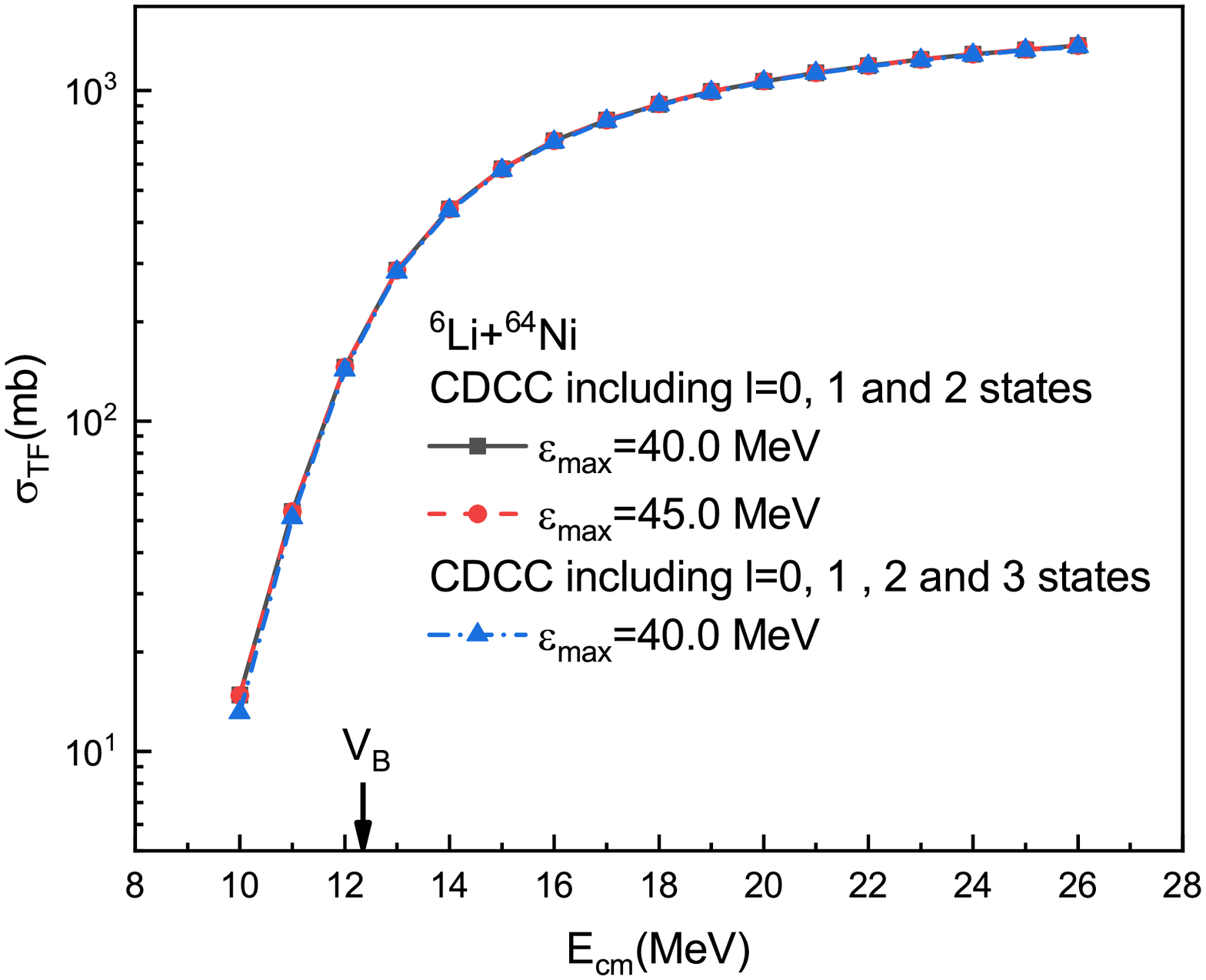}
  \caption{(Color online) Calculated TF cross sections for $^6$Li+$^{64}$Ni system with different conditions. The arrow indicates the Coulomb barrier. See text for details.}
  \label{fig-TF-6+64-convergence}
\end{figure}

In Fig. \ref{fig-TF-6+64-convergence}, an example of the convergence of TF cross sections calculations is shown for the $^6$Li+$^{64}$Ni system at energies around the Coulomb barrier. In these calculations, $W_0$=50.0 MeV. The CDCC calculations with $\varepsilon_{\max}$=40.0 and 45.0 MeV are almost the same when $l$=0, 1 and 2 states are included. The inclusion of the states with higher $l$ has an invisible effect on TF cross sections. Therefore, the convergence of TF cross section calculations in the present work is ensured.

\section{Calculated results and analysis} \label{sec-4}

In this section, we study the continuum coupling effect on total and complete fusion. The CDCC calculations are performed with $W_0$=50.0 MeV and $l$=0, 1 and 2 continuum states. Four types of calculations are performed:

(1) No continuum states are included, i.e., single channel calculations;

(2) Continuum states below $\varepsilon_{max}$=6.0 MeV are included, which include the resonance states;

(3) Continuum states below $\varepsilon_{max}$=$\varepsilon_{th}$ are included, closed channels are ignored;

(4) Continuum states below $\varepsilon_{max}$=40.0 MeV are included.

It should be emphasised that the second type of calculation is not made for $^6$Li+$^{28}$Si system at some low incident energies to avoid the inclusion of closed channels. The fourth type of calculation is not performed for $^6$Li+$^{209}$Bi system at a few high energies, where $\varepsilon_{th}$ is larger than 40.0 MeV.

Although the second and third types of calculations can not provide converged TF cross sections, they can be applied to analyse the coupling effect of the continuum states in different energy regions. In the present work, we mainly focus on the coupling effect from the continuum states above the resonance energy regions ($\varepsilon \geq$ 6.0 MeV). The coupling effects of open and closed channels can be distinguished by comparing the results of the third and fourth types of calculations.

\subsection{Continuum coupling effect on total fusion}\label{sec-4-1}

\begin{figure}[tbp]
  \centering
  \includegraphics[width=\columnwidth]{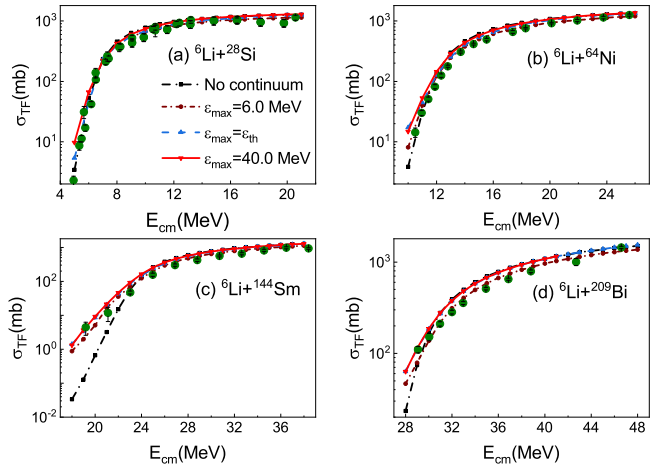}
  \caption{(Color online) Calculated TF cross sections for $^6$Li+$^{28}$Si, $^{64}$Ni, $^{144}$Sm and $^{209}$Bi systems shown in logarithmic scale. Experimental data are taken from Refs. \cite{Pakou2009,Sinha2010,Hugi1981,Shaikh2014,Sinha2017a,Rath2009,Rath2009a,Dasgupta2002,Dasgupta2004}. See text for details.}
  \label{fig-TF-all-semilogy}
\end{figure}

\begin{figure}[htbp]
  \centering
  \includegraphics[width=\columnwidth]{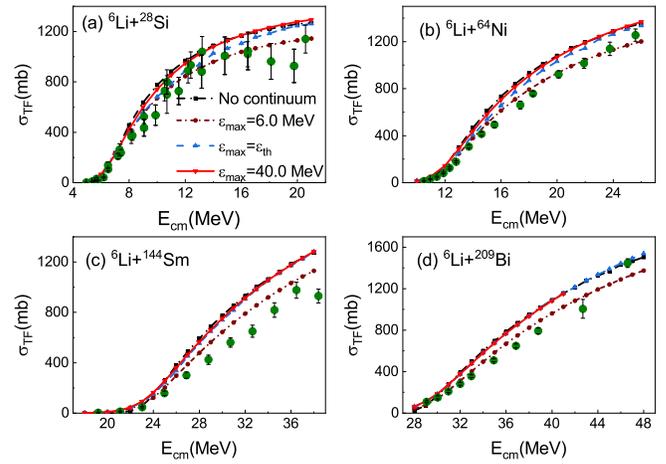}
  \caption{(Color online) Same as Fig. \ref{fig-TF-all-semilogy} but shown in linear scale.}
  \label{fig-TF-all-linear}
\end{figure}

Figures \ref{fig-TF-all-semilogy} and \ref{fig-TF-all-linear} show the calculated TF cross sections for $^6$Li+$^{28}$Si, $^{64}$Ni, $^{144}$Sm and $^{209}$Bi systems, and their comparison with the experimental data \cite{Pakou2009,Sinha2010,Hugi1981,Sinha2017a,Shaikh2014,Rath2009,Rath2009a,Dasgupta2002,Dasgupta2004}. The two figures are shown in logarithmic and linear scale, convenient to see the TF cross sections at energies below and above the Coulomb barrier respectively.

In general, the agreement between the converged CDCC results and the measured value is reasonably good in the sub-barrier region, but the converged CDCC results overestimate the experimental data at energies above the Coulomb barrier for $^{64}$Ni, $^{144}$Sm and $^{209}$Bi targets. This overestimation may be resulted by the following reasons:

(1) The adopted nuclear potentials for $\alpha$ and $d$ with targets (SPP2) are the systematic research results for various elastic scattering experimental data and it is expected that there are some disagreements for specific reaction systems.

(2) In the present work, only the $^6$Li$\rightarrow \alpha+d$ channel is taken into account. However, there are other important reaction channels, such as $n-$transfer, which can influence the TF cross sections.

(3) Rath et al. \cite{Rath2009a} measured the incomplete fusion cross section for $^6$Li+$^{144}$Sm by summing the detected cross sections for $d$- and $\alpha$-capture, in which not all possible evaporation residue channels were included. Therefore, the values of the ICF and TF(=CF+ICF) can be considered as the lower limit. Similarly, Dasgupta et al. \cite{Dasgupta2002,Dasgupta2004} measured the fusion cross sections for $^6$Li+$^{209}$Bi system by detecting $\alpha$ particles emitted by the evaporation residues of the compound nuclei. However, they missed the contribution from $^{209}$Po, which could not be measured due to its long half-life (about 200 yr). According to the statistical model estimation \cite{Dasgupta2004}, the contribution from $^{209}$Po was excepted to be significant at the high energies.

For all the four collision systems, it can be seen in Figs. \ref{fig-TF-all-semilogy} and \ref{fig-TF-all-linear} that the results of the CDCC calculations become close to the single channel results in the above-barrier region by varying $\varepsilon_{max}$ from 6.0 MeV to 40.0 MeV. In the sub-barrier region, the calculated TF cross sections are enhanced by increasing $\varepsilon_{max}$. To determine the continuum coupling effect on TF cross section, we define
\begin{equation}\label{e-TFY}
  \Gamma _i=\frac{\sigma _{TF}\left( i \right)}{\sigma _{TF}\left( 1 \right)}-1,i=2,3,4.
\end{equation}
The $\sigma_{TF}(i) (i=1,2,3,4)$ denotes the TF cross section obtained by the $i$-th type of calculation as listed in the beginning of Sec. \ref{sec-4}.

\begin{figure}[tbp]
  \centering
  \includegraphics[width=\columnwidth]{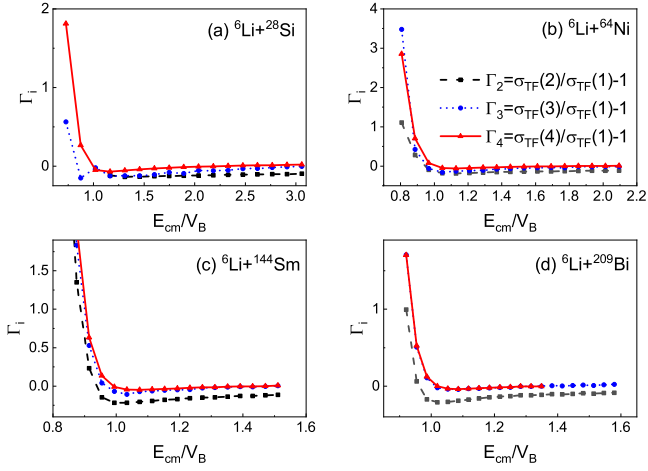}
  \caption{(Color online) $\Gamma_i$ ($i$=2, 3 and 4) for $^6$Li+$^{28}$Si, $^{64}$Ni, $^{144}$Sm and $^{209}$Bi systems.}
  \label{fig-TF-gamma}
\end{figure}

Fig. \ref{fig-TF-gamma} presents the $\Gamma_i$-values ($i$=2, 3 and 4) for $^6$Li+$^{28}$Si, $^{64}$Ni, $^{144}$Sm and $^{209}$Bi systems. Compared with the TF cross sections obtained without continuum states coupling, there is a 10-15$\%$ suppression at energies above the Coulomb barrier for these reaction systems when $\varepsilon_{max}$=6.0 MeV (See the results of $\Gamma_2$). However, this suppression is much reduced for $^6$Li+$^{28}$Si and eliminated for $^6$Li+$^{64}$Ni, $^{144}$Sm and $^{209}$Bi when the converged result is obtained with $\varepsilon_{max}$=40.0 MeV (See the results of $\Gamma_4$). In the sub-barrier region, the theoretical TF cross sections are further enhanced by increasing $\varepsilon_{max}$. A detailed discussion is given below.

For the $^6$Li+$^{28}$Si system, with the $\varepsilon_{max}$ increasing from 6.0 MeV to $\varepsilon_{th}$, the suppression remains nearly unchanged in the region 1.0 $\leq E_{cm} / V_B \leq $ 1.8 and is moderately reduced in 1.8 $\leq E_{cm} / V_B \leq $ 2.5 (See the results of $\Gamma_3$). Only when the closed channels are taken into CDCC calculations can the suppression effect be negligible. In higher incident energies, the TF cross sections obtained with  $\varepsilon_{max}$=$\varepsilon_{th}$ and 40.0 MeV are almost the same, indicating that the coupling effect of closed channels can be ignored in this high energy region. In the sub-barrier region, a much stronger enhancement is provided by the calculation with $\varepsilon_{max}$=40.0 MeV, where the coupling effect of closed channels is completely taken into account.

For the medium mass targets $^{64}$Ni and $^{144}$Sm, the calculated results obtained with $\varepsilon_{max}$=$\varepsilon_{th}$ are slightly smaller than the single channel results in a very narrow region
around $E_{cm}$. But the converged CDCC calculation, which includes the closed channels, has only an enhanced effect on $\sigma_{TF}$ at energies below the Coulomb barrier. When $E_{cm} >$  1.1 $V_B$, the values obtained with $\varepsilon_{max}$=$\varepsilon_{th}$ and 40.0 MeV are almost the same as the single channel results. In the energy region $E_{cm} <$  0.9 $V_B$, the CDCC calculations with $\varepsilon_{max}$=$\varepsilon_{th}$ and 40.0 MeV provide a similar enhancement effect on TF cross section, which is stronger than that given by the calculation with $\varepsilon_{max}$=6.0 MeV.

As the Coulomb barrier for $^6$Li+$^{209}$Bi system is quite high, the results obtained with $\varepsilon_{max}$=40.0 MeV  are almost the same as those obtained with $\varepsilon_{max}$=$\varepsilon_{th}$. Their results are approximately equal to the results obtained without continuum states coupling at energies above the Coulomb barrier. In the sub-barrier region, the TF cross sections are further enhanced by taking higher energy continuum states ($\varepsilon \geq$ 6.0 MeV) into CDCC calculations.

The above results seem to suggest that the coupling effects from low-energy continuum states (0$\leq \varepsilon \leq$ 6.0 MeV) and higher-energy continuum states (6.0 $\leq \varepsilon \leq$ 40.0 MeV) add destructively for the total fusion cross sections at energies above the Coulomb barrier. In the end, the converged CDCC results provide nearly the same total fusion reactions as single-channel calculations. This is an interesting observation.

\subsection{Continuum coupling effect on complete fusion}\label{sec-4-2}

Based on the converged TF cross sections and the experimental data for complete fusion \cite{Sinha2017a,Shaikh2014,Rath2009,Rath2009a,Dasgupta2002,Dasgupta2004}, we apply the sum-rule model to extract the cut-off angular momentum $J_c$ and CF cross section $\sigma_{CF}$, as shown in Eq. (\ref{e-CF}).

In the early researches \cite{Wilczynski1973386,Wilczynski1980,Wilczynski1982109}, the sum-rule model was applied to the reactions at incident energies well above the Coulomb barrier. In these cases, CF and ICF processes can be well separated by a critical angular momentum, $J_{crit}$, which is nearly independent of the incident energy. Recently, Mukeru et al. combined this model with CDCC method to study the fusion of weakly bound nuclei at energies around the Coulomb barrier, such as $^{8}$Li+$^{208}$Pb \cite{Mukeru2020121700} and $^{9}$Be+$^{124}$Sn, $^{144}$Sm and $^{208}$Pb \cite{Mukeru2021}. The angular momentum to separate the CF and ICF processes, $J_c$, was found to be incident-energy-dependent.

\begin{figure}[tbp]
  \centering
  \includegraphics[width=\columnwidth]{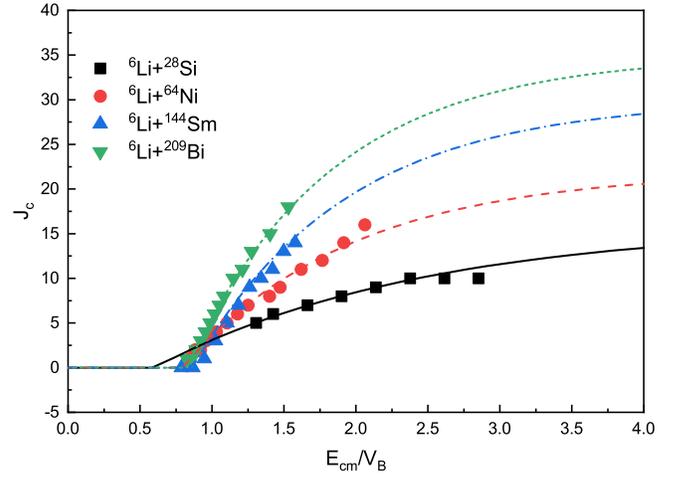}
  \caption{(Color online) The extracted cut-off angular momentum $J_c$ for $^6$Li+$^{28}$Si (squares), $^{64}$Ni (circles), $^{144}$Sm (upper triangular) and $^{209}$Bi (lower triangular) systems. The corresponding fitting curves are represented by the solid, dashed, dashed-dotted and short-dashed lines respectively.}
  \label{fig-Jc}
\end{figure}

\begin{table}[tbp]
\caption{The values of $C_0$, $C_1$ and $J_{crit}$.}
\label{table-Jc}
\begin{ruledtabular}
\begin{tabular}{cccc}
 system & $C_0$ & $C_1$ & $J_{crit}$ \\
\hline
 $^6$Li+$^{28}$Si    & 0.54 & 0.60 & 16 \\
 $^6$Li+$^{64}$Ni    & 0.86 & 0.80 & 22 \\
 $^6$Li+$^{144}$Sm   & 0.94 & 0.86 & 30 \\
 $^6$Li+$^{209}$Bi   & 1.00 & 0.82 & 35 \\
\end{tabular}
\end{ruledtabular}
\end{table}

Following the method of Mukeru et al. \cite{Mukeru2020121700,Mukeru2021}, $J_c$ can be linked to the critical angular momentum $J_{crit}$ with an analytical expression
\begin{equation}\label{e-Jc-Jcrit}
  J_c=\left\{ 1-\exp \left[ -C_0\left( \frac{E_{cm}}{V_B}-C_1 \right) \right] \right\} J_{crit}.
\end{equation}
$J_{crit}$ is calculated with the formalism in Ref. \cite{Wilczynski1973386}. $C_0$ and $C_1$ are obtained by fitting $J_c$. Their values are listed in Table \ref{table-Jc}. The extracted $J_c$ and the fitting curves are plotted in Fig. \ref{fig-Jc}. The fitted value is set to 0 when it is negative. As the energy decreases, it can be seen obviously that the fitting curve for $^6$Li+$^{28}$Si returns to zero at $E_{cm}/V_B \approx$0.6 while those for other systems return to 0 at $E_{cm}/V_B \approx$0.8. As there is no available CF data for $^6$Li+$^{28}$Si system in the sub-barrier energy region, this difference is expected to be explained with further experimental and theoretical studies.

\begin{figure}[tbp]
  \centering
  \includegraphics[width=\columnwidth]{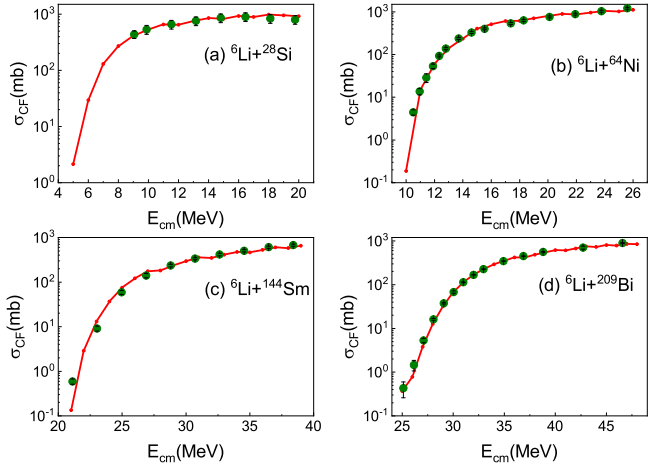}
  \caption{(Color online) The CF cross sections extracted though the sum-rule model for $^6$Li+$^{28}$Si, $^{64}$Ni, $^{144}$Sm and $^{209}$Bi systems, compared with the experimental data \cite{Sinha2017a,Shaikh2014,Rath2009,Rath2009a,Dasgupta2002,Dasgupta2004}.}
  \label{fig-CF-all}
\end{figure}

The calculated CF cross sections are compared with experimental data in Fig. \ref{fig-CF-all}. Good agreement is obtained between theory and experiment. To analyse the continuum coupling effect on CF process, we pay close attention to the partial wave fusion cross sections $\sigma_{J}$ in the partial waves below the cut-off angular momentum $J_c$, at energies below, near, above and well above the Coulomb barrier. In addition, it is practical to apply $J_c$ for the single channel calculation to extract the cross section $\sigma_{CF}^{1 ch}$, which can be regarded as the CF cross section without continuum coupling. It can be compared with $\sigma_{CF}$ to determine the continuum effect on complete fusion quantitatively.

\begin{figure}[htbp]
  \centering
  \includegraphics[width=\columnwidth]{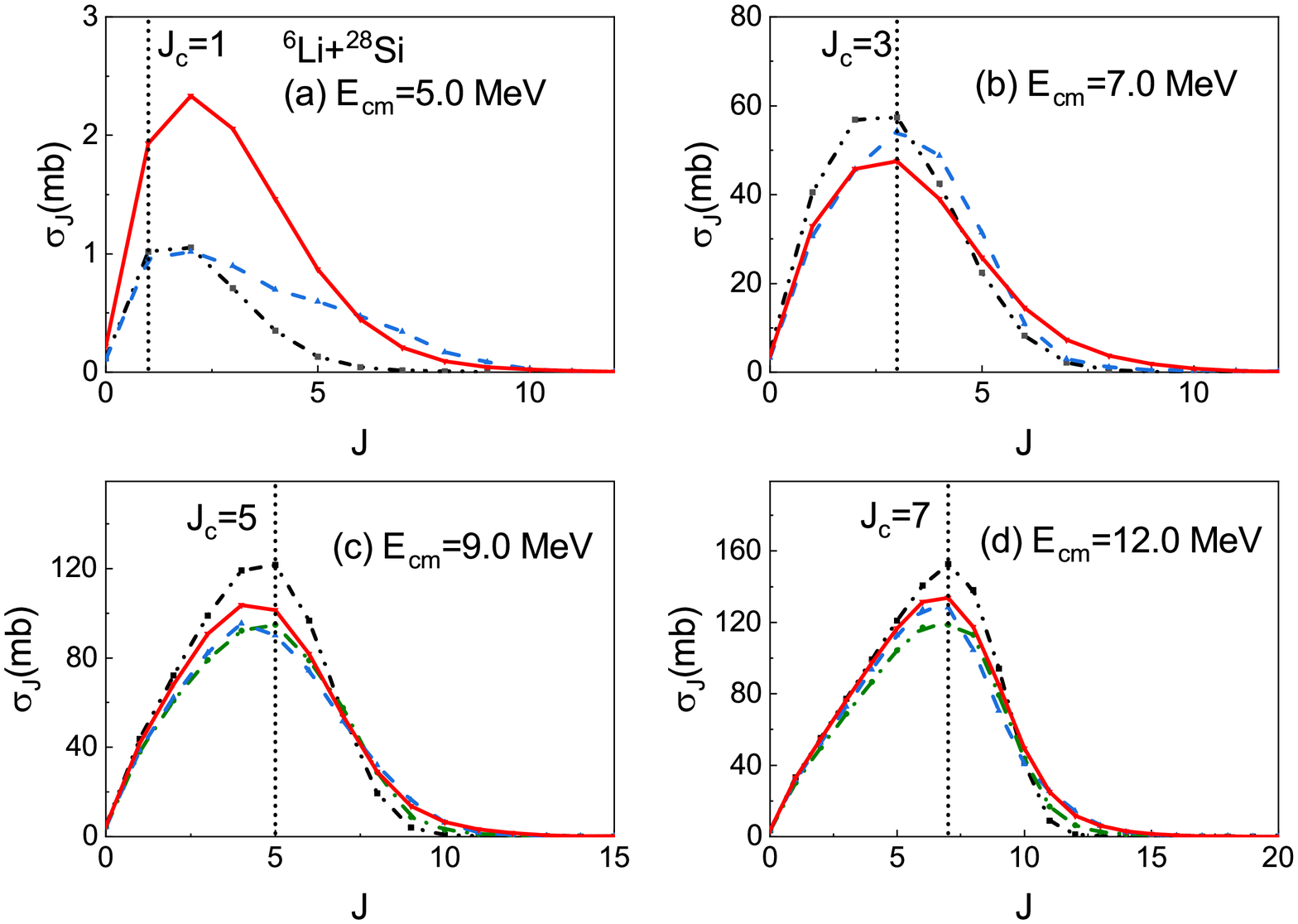}
  \caption{(Color online) Partial wave fusion cross section $\sigma_{J}$ for $^{6}$Li+$^{28}$Si at $E_{cm}$=5.0, 7.0, 9.0 and 12.0 MeV. The four energy points correspond to the energies below, near, above and well above the Coulomb barrier respectively. The calculated results without continuum coupling and those with $\varepsilon_{max}$=6.0 MeV, $\varepsilon_{th}$ and 40.0 MeV are plotted in black dashed-double-dotted, green dashed-dotted, blue dashed and red solid lines respectively. The short dotted lines represent the cut-off angular momentum $J_c$.}
  \label{fig-CF-J-28Si}
\end{figure}

Fig. \ref{fig-CF-J-28Si} shows the partial wave fusion cross sections $\sigma_J$ for $^{6}$Li+$^{28}$Si system at $E_{cm}$=5.0, 7.0, 9.0 and 12.0 MeV. At the first energy point which is below the Coulomb barrier, the $\sigma_{J}$ of the single-channel calculation and the CDCC calculation with $\varepsilon_{max}$=$\varepsilon_{th}$ at $J\leq J_{c}$ are almost the same. It suggests that in this case, the continuum states of open channels have little coupling effect on complete fusion, although the $3^+$ and 2$^+$ resonance states have been taken into account. Only when the closed channels are taken into calculations will the $\sigma_J$ in lower angular momenta be significantly enhanced. At $E_{cm}$=9.0 and 12.0 MeV, the converged $\sigma_{J}$ in $J \leq J_c$ is slightly larger than those calculated with $\varepsilon_{max}$=6.0 MeV and $\varepsilon_{th}$. But they are all visibly smaller than the single channel results in $J_c-2 \leq J \leq J_c$. Because of the continuum coupling, the $\sigma_{CF}$ in the above-barrier region is smaller than $\sigma_{CF}^{1 ch}$ about 5-20$\%$, which is consistent with the suppression factor of 15$\%$ given by Mandira and Lubian \cite{Sinha2017a}.

\begin{figure}[htbp]
  \centering
  \includegraphics[width=\columnwidth]{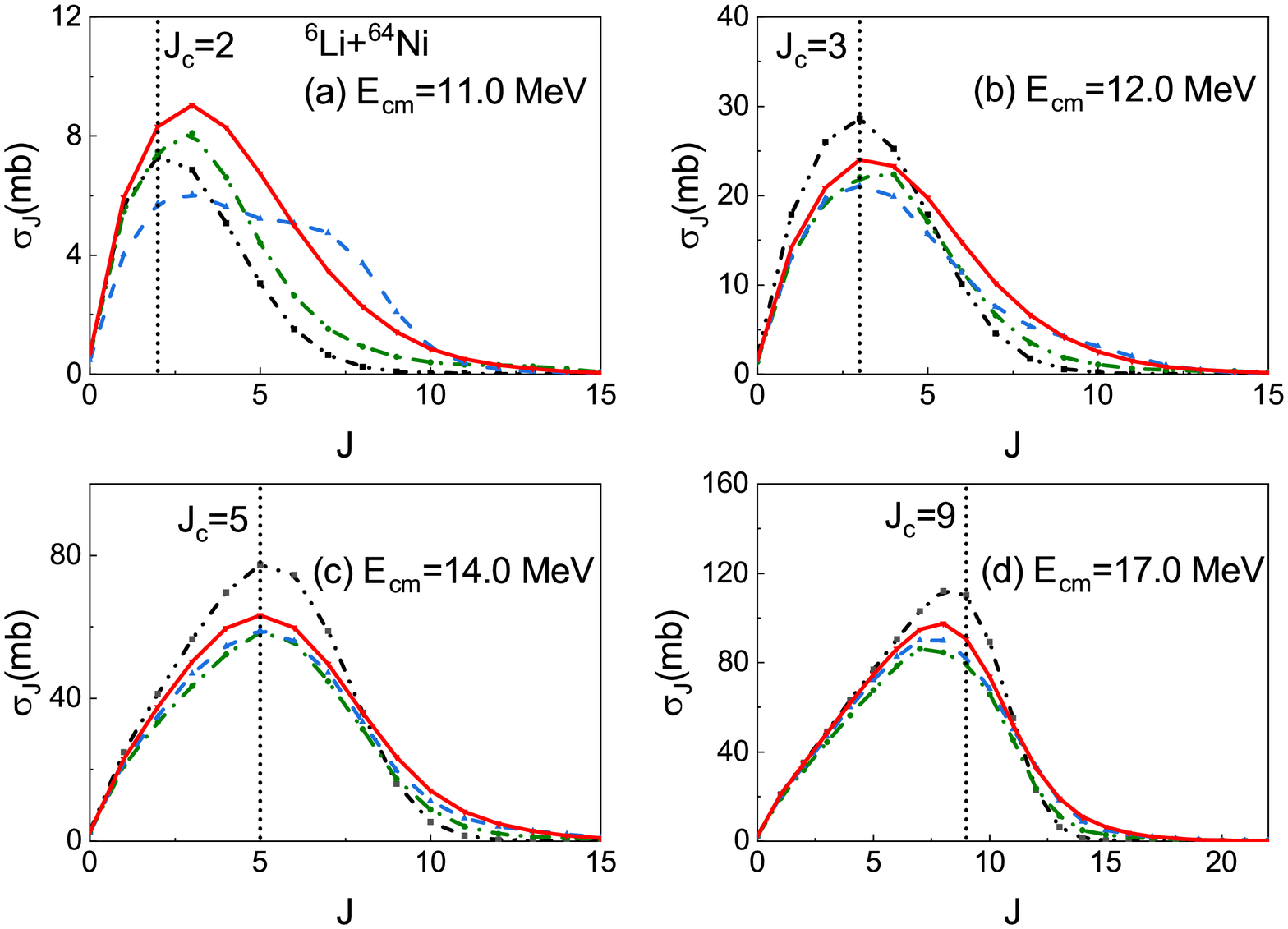}
  \caption{(Color online) Same as Fig. \ref{fig-CF-J-28Si} but for $^{6}$Li+$^{64}$Ni at $E_{cm}$=11.0, 12.0, 14.0 and 17.0 MeV.}
  \label{fig-CF-J-64Ni}
\end{figure}

The partial wave fusion cross sections for $^6$Li+$^{64}$Ni system are presented in Fig. \ref{fig-CF-J-64Ni}. There is a noticeable issue shown in Fig. \ref{fig-CF-J-64Ni}(a). At the energy (11.0 MeV) below the Coulomb barrier, the CDCC calculated $\sigma_{J}$ in $J \leq J_c$ decreases firstly when $\varepsilon_{max}$ increases from 6.0 MeV to $\varepsilon_{th}$ and it then increases by including closed channels. Eventually, the converged $\sigma_{J}$ is slightly larger than the single channel result in $J \leq J_c$. As the $\sigma_J$ in $J \leq J_c$ obtained with the single channel calculation and the CDCC calculation with $\varepsilon_{max}$=6.0 MeV are almost the same, it can be concluded that the enhancement on $\sigma_{CF}$ mainly comes from the continuum states above 6.0 MeV and the closed channels have a fundamental coupling effect. At $E_{cm}$=12.0, 14.0 and 17.0 MeV, the $\sigma_J$ in $J \leq J_c$ obtained with $\varepsilon_{max}$=40.0 MeV are moderately larger than the results of other two kinds of CDCC calculations, while they are all smaller than the single channel result in $J_c-2 \leq J \leq J_c$ observably. By comparing $\sigma_{CF}^{1 ch}$ and $\sigma_{CF}$, a 13-18$\%$ suppression factor is given at energies above the Coulomb barrier, in good agreement with the result of 13$\pm$7$\%$ in Ref. \cite{Shaikh2014}.

\begin{figure}[tbp]
  \centering
  \includegraphics[width=\columnwidth]{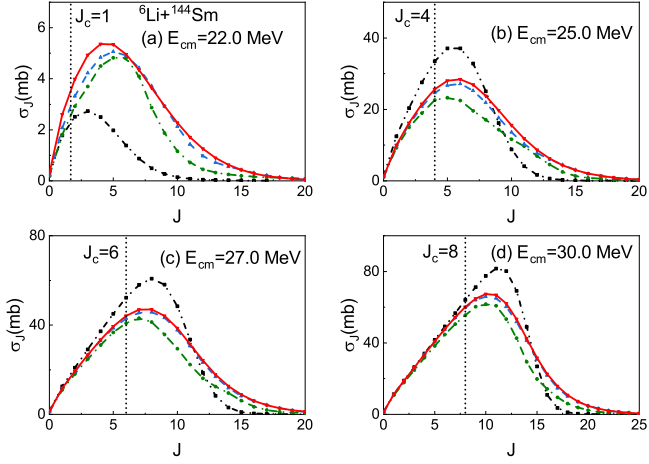}
  \caption{(Color online) Same as Fig. \ref{fig-CF-J-28Si} but for $^{6}$Li+$^{144}$Sm at $E_{cm}$=22.0, 25.0, 27.0 and 30.0 MeV.}
  \label{fig-CF-J-144Sm}
\end{figure}

In Fig. \ref{fig-CF-J-144Sm}, the partial wave fusion cross sections $\sigma_{J}$ for $^{6}$Li+$^{144}$Sm system are plotted at $E_{cm}$=22.0, 25.0, 27.0 and 30.0 MeV. As $\varepsilon_{max}$ increases from 6.0 MeV to 40.0 MeV, the $\sigma_{J}$ in  $J \leq J_c$ is slightly enlarged for all four energy points. The $\sigma_{CF}$ is larger than the $\sigma_{CF}^{1 ch}$ at $E_{cm}$=22.0, while it becomes smaller about 5-25$\%$ at other three energies because of the reduction of $\sigma_{J}$ in $J_c-2 \leq J \leq J_{c}$. It is reasonably consistent with the suppression of 32$\pm$5$\%$ in Ref. \cite{Rath2009}.

\begin{figure}[htbp]
  \centering
  \includegraphics[width=\columnwidth]{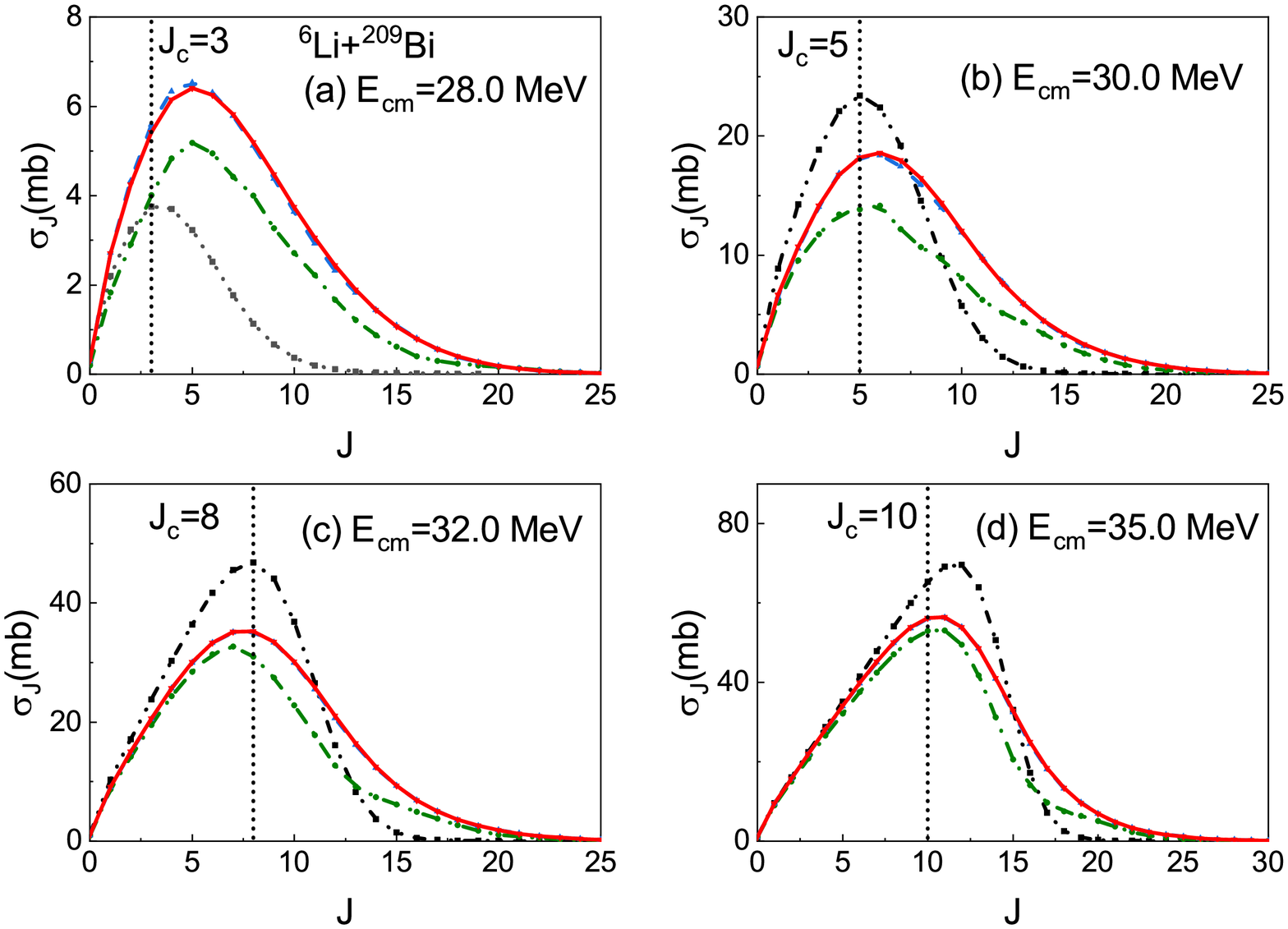}
  \caption{(Color online) Same as Fig. \ref{fig-CF-J-28Si} but for $^{6}$Li+$^{209}$Bi at $E_{cm}$=28.0, 30.0, 32.0 and 35.0 MeV.}
  \label{fig-CF-J-209Bi}
\end{figure}

The partial wave fusion cross sections $\sigma_{J}$ for $^{6}$Li+$^{209}$Bi are presented in Fig. \ref{fig-CF-J-209Bi} at $E_{cm}$=28.0, 30.0, 32.0 and 35.0 MeV. The results obtained with $\varepsilon_{max}$=$\varepsilon_{th}$ and 40.0 MeV are almost the same in all partial waves, suggesting that the closed channels have hardly any coupling effect on this reaction system. On the other hand, the $\sigma_{J}$ in $J \leq J_c$ is visibly enlarged at $E_{cm}$=28.0 and 30.0 MeV by increasing $\varepsilon_{max}$ from 6.0 MeV to $\varepsilon_{th}$, showing the obvious coupling effect of the continuum states above the resonance energy region. With the same improvement of $\varepsilon_{max}$, there is a slight enhancement on $\sigma_{J}$ in $J \leq J_c$ at $E_{cm}$=32.0 and 35.0 MeV. Similarly with other reaction systems, $\sigma_{CF}$ is higher than $\sigma_{CF}^{1ch}$ at $E_{cm}$=28.0 MeV but becomes lower about 10-35$\%$ at higher energies, which agrees with the suppression of 36$\pm$3$\%$ provided by Dasgupta et al. \cite{Dasgupta2004}.

In addition, for all four reaction systems, we found that:

(1) At energies below the Coulomb barrier, the $\sigma_J$ obtained by the converged CDCC calculation is larger than that obtained by single channel calculation at each $J$. Based on the sum-rule model, it indicates that continuum coupling increases both the complete fusion and incomplete fusion cross sections in the sub-barrier energy region.

(2) At energies above the Coulomb barrier, the single channel calculated $\sigma_J$ is larger than that obtained by the converged CDCC calculation in angular momentum region around $J_c$ but it becomes smaller at relatively higher $J$. Finally, the single channel calculation and the converged CDCC calculation provide close values for the total fusion cross section, as shown in Fig. \ref{fig-TF-all-linear}. It can be deduced that continuum coupling reduces and increases the probabilities of CF and ICF processes respectively in the above-barrier energy region.

As our present work overestimates the TF cross section at energies above the Coulomb barrier, we do not further investigate the continuum coupling effect on incomplete fusion in this paper. A comprehensive study of the continuum coupling effect on $^6$Li fusion will be done in the future with a good description of CF, ICF and TF processes.

\section{Summary and conclusion} \label{sec-5}

CDCC calculations have been presented for the fusion reactions of weakly bound projectile $^6$Li with $^{28}$Si, $^{64}$Ni, $^{144}$Sm and $^{209}$Bi targets. The inclusion of continuum states up to 40 MeV was found necessary for the convergence of the total fusion cross sections, which means that the inclusion of closed channels in CDCC calculations is necessary for $^6$Li fusion with light and medium mass targets, such as $^{28}$Si, $^{64}$Ni and $^{144}$Sm, at incident energies around the Coulomb barriers. Reasonable agreement between the calculated TF cross section and experimental data is obtained.

At energies above the Coulomb barrier, it is found that the continuum coupling effects induced by low-energy (0$\leq \varepsilon \leq$ 6.0 MeV) and higher-energy (6.0 $\leq \varepsilon \leq$ 40.0 MeV) continuum states are different for TF cross sections: the former reduces the TF cross sections by around 10-15$\%$, but the latter increases the TF cross sections. Therefore the final converged results are nearly the same as the results of single channel calculations, in which no continuum coupling effects were taken into account.

In the sub-barrier region, the calculated TF cross section is further enhanced by including the higher-energy continuum states.

The Sum-rule model has been adopted to extract the complete fusion cross section as well as the critical angular momentum $J_c$. The coupling effect of continuum states can not be ignored for the complete fusion process, especially at energies below the Coulomb barrier. In particular, the enhancement effect on the complete fusion cross section in the sub-barrier region is dominated by closed channels for the $^6$Li+$^{28}$Si, $^{64}$Ni and $^{144}$Sm systems. At energies above the Coulomb barrier, the CDCC calculated complete fusion cross section can be slightly enlarged by taking the higher-energy continuum states into account, while the converged result is still smaller than the single channel result. In general, the extracted suppression factors for complete fusion in the above-barrier region are consistent with previous studies for these reaction systems. It is found that the suppression mainly occurs in the angular momentum region $J_c-2 \leq J \leq J_c$, which is independent of the fusion system. We believe that this independence calls for full-quantum investigations.

Finally, as the coupling effect of the higher-energy continuum states (including both open and closed channels) is remarkable for $^6$Li induced fusion reactions, it is of interest to study their effect on the fusion reactions induced by other weakly bound nuclei, such as $^7$Li and $^9$Be. Related research is in progress.

\begin{acknowledgments}
 This work was financially supported by the National Natural Science Foundation of China (Grant No. U2067205).
\end{acknowledgments}

\bibliography{thesis}

\providecommand{\noopsort}[1]{}\providecommand{\singleletter}[1]{#1}%
\begin{thebibliography}{56}%
\makeatletter
\providecommand \@ifxundefined [1]{%
 \@ifx{#1\undefined}
}%
\providecommand \@ifnum [1]{%
 \ifnum #1\expandafter \@firstoftwo
 \else \expandafter \@secondoftwo
 \fi
}%
\providecommand \@ifx [1]{%
 \ifx #1\expandafter \@firstoftwo
 \else \expandafter \@secondoftwo
 \fi
}%
\providecommand \natexlab [1]{#1}%
\providecommand \enquote  [1]{``#1''}%
\providecommand \bibnamefont  [1]{#1}%
\providecommand \bibfnamefont [1]{#1}%
\providecommand \citenamefont [1]{#1}%
\providecommand \href@noop [0]{\@secondoftwo}%
\providecommand \href [0]{\begingroup \@sanitize@url \@href}%
\providecommand \@href[1]{\@@startlink{#1}\@@href}%
\providecommand \@@href[1]{\endgroup#1\@@endlink}%
\providecommand \@sanitize@url [0]{\catcode `\\12\catcode `\$12\catcode
  `\&12\catcode `\#12\catcode `\^12\catcode `\_12\catcode `\%12\relax}%
\providecommand \@@startlink[1]{}%
\providecommand \@@endlink[0]{}%
\providecommand \url  [0]{\begingroup\@sanitize@url \@url }%
\providecommand \@url [1]{\endgroup\@href {#1}{\urlprefix }}%
\providecommand \urlprefix  [0]{URL }%
\providecommand \Eprint [0]{\href }%
\providecommand \doibase [0]{https://doi.org/}%
\providecommand \selectlanguage [0]{\@gobble}%
\providecommand \bibinfo  [0]{\@secondoftwo}%
\providecommand \bibfield  [0]{\@secondoftwo}%
\providecommand \translation [1]{[#1]}%
\providecommand \BibitemOpen [0]{}%
\providecommand \bibitemStop [0]{}%
\providecommand \bibitemNoStop [0]{.\EOS\space}%
\providecommand \EOS [0]{\spacefactor3000\relax}%
\providecommand \BibitemShut  [1]{\csname bibitem#1\endcsname}%
\let\auto@bib@innerbib\@empty
%</preamble>
\bibitem [{\citenamefont {Canto}\ \emph {et~al.}(2006)\citenamefont {Canto},
  \citenamefont {Gomes}, \citenamefont {Donangelo},\ and\ \citenamefont
  {Hussein}}]{Canto20061}%
  \BibitemOpen
  \bibfield  {author} {\bibinfo {author} {\bibfnamefont {L.}~\bibnamefont
  {Canto}}, \bibinfo {author} {\bibfnamefont {P.}~\bibnamefont {Gomes}},
  \bibinfo {author} {\bibfnamefont {R.}~\bibnamefont {Donangelo}},\ and\
  \bibinfo {author} {\bibfnamefont {M.}~\bibnamefont {Hussein}},\ }\bibfield
  {title} {\bibinfo {title} {Fusion and breakup of weakly bound nuclei},\
  }\href {https://doi.org/https://doi.org/10.1016/j.physrep.2005.10.006}
  {\bibfield  {journal} {\bibinfo  {journal} {Phys. Rep.}\ }\textbf {\bibinfo
  {volume} {424}},\ \bibinfo {pages} {1} (\bibinfo {year} {2006})}\BibitemShut
  {NoStop}%
\bibitem [{\citenamefont {Keeley}\ \emph {et~al.}(2009)\citenamefont {Keeley},
  \citenamefont {Alamanos}, \citenamefont {Kemper},\ and\ \citenamefont
  {Rusek}}]{Keeley2009396}%
  \BibitemOpen
  \bibfield  {author} {\bibinfo {author} {\bibfnamefont {N.}~\bibnamefont
  {Keeley}}, \bibinfo {author} {\bibfnamefont {N.}~\bibnamefont {Alamanos}},
  \bibinfo {author} {\bibfnamefont {K.}~\bibnamefont {Kemper}},\ and\ \bibinfo
  {author} {\bibfnamefont {K.}~\bibnamefont {Rusek}},\ }\bibfield  {title}
  {\bibinfo {title} {Elastic scattering and reactions of light exotic beams},\
  }\href {https://doi.org/https://doi.org/10.1016/j.ppnp.2009.05.003}
  {\bibfield  {journal} {\bibinfo  {journal} {Prog. Part. Nucl. Phys.}\
  }\textbf {\bibinfo {volume} {63}},\ \bibinfo {pages} {396} (\bibinfo {year}
  {2009})}\BibitemShut {NoStop}%
\bibitem [{\citenamefont {Hagino}\ and\ \citenamefont
  {Takigawa}(2012)}]{Hagino2012}%
  \BibitemOpen
  \bibfield  {author} {\bibinfo {author} {\bibfnamefont {K.}~\bibnamefont
  {Hagino}}\ and\ \bibinfo {author} {\bibfnamefont {N.}~\bibnamefont
  {Takigawa}},\ }\bibfield  {title} {\bibinfo {title} {{Subbarrier Fusion
  Reactions and Many-Particle Quantum Tunneling}},\ }\href
  {https://doi.org/10.1143/PTP.128.1061} {\bibfield  {journal} {\bibinfo
  {journal} {Phys. Rep.}\ }\textbf {\bibinfo {volume} {128}},\ \bibinfo {pages}
  {1061} (\bibinfo {year} {2012})}\BibitemShut {NoStop}%
\bibitem [{\citenamefont {Canto}\ \emph {et~al.}(2015)\citenamefont {Canto},
  \citenamefont {Gomes}, \citenamefont {Donangelo}, \citenamefont {Lubian},\
  and\ \citenamefont {Hussein}}]{Canto20151}%
  \BibitemOpen
  \bibfield  {author} {\bibinfo {author} {\bibfnamefont {L.}~\bibnamefont
  {Canto}}, \bibinfo {author} {\bibfnamefont {P.}~\bibnamefont {Gomes}},
  \bibinfo {author} {\bibfnamefont {R.}~\bibnamefont {Donangelo}}, \bibinfo
  {author} {\bibfnamefont {J.}~\bibnamefont {Lubian}},\ and\ \bibinfo {author}
  {\bibfnamefont {M.}~\bibnamefont {Hussein}},\ }\bibfield  {title} {\bibinfo
  {title} {Recent developments in fusion and direct reactions with weakly bound
  nuclei},\ }\href
  {https://doi.org/https://doi.org/10.1016/j.physrep.2015.08.001} {\bibfield
  {journal} {\bibinfo  {journal} {Phys. Rep.}\ }\textbf {\bibinfo {volume}
  {596}},\ \bibinfo {pages} {1} (\bibinfo {year} {2015})}\BibitemShut {NoStop}%
\bibitem [{\citenamefont {Lei}\ and\ \citenamefont {Moro}(2019)}]{Lei2019}%
  \BibitemOpen
  \bibfield  {author} {\bibinfo {author} {\bibfnamefont {J.}~\bibnamefont
  {Lei}}\ and\ \bibinfo {author} {\bibfnamefont {A.~M.}\ \bibnamefont {Moro}},\
  }\bibfield  {title} {\bibinfo {title} {Puzzle of complete fusion suppression
  in weakly bound nuclei: A trojan horse effect?},\ }\href
  {https://doi.org/10.1103/PhysRevLett.122.042503} {\bibfield  {journal}
  {\bibinfo  {journal} {Phys. Rev. Lett.}\ }\textbf {\bibinfo {volume} {122}},\
  \bibinfo {pages} {042503} (\bibinfo {year} {2019})}\BibitemShut {NoStop}%
\bibitem [{\citenamefont {Beck}\ \emph {et~al.}(2003)\citenamefont {Beck},
  \citenamefont {Souza}, \citenamefont {Rowley}, \citenamefont {Sanders},
  \citenamefont {Aissaoui}, \citenamefont {Alonso}, \citenamefont {Bednarczyk},
  \citenamefont {Carlin}, \citenamefont {Courtin}, \citenamefont {Diaz-Torres},
  \citenamefont {Dummer}, \citenamefont {Haas}, \citenamefont {Hachem},
  \citenamefont {Hagino}, \citenamefont {Hoellinger}, \citenamefont {Janssens},
  \citenamefont {Kintz}, \citenamefont {Liguori~Neto}, \citenamefont {Martin},
  \citenamefont {Moura}, \citenamefont {Munhoz}, \citenamefont {Papka},
  \citenamefont {Rousseau}, \citenamefont {S\`anchez~i Zafra}, \citenamefont
  {St\'ezowski}, \citenamefont {Suaide}, \citenamefont {Szanto}, \citenamefont
  {Szanto~de Toledo}, \citenamefont {Szilner},\ and\ \citenamefont
  {Takahashi}}]{Beck2003}%
  \BibitemOpen
  \bibfield  {author} {\bibinfo {author} {\bibfnamefont {C.}~\bibnamefont
  {Beck}}, \bibinfo {author} {\bibfnamefont {F.~A.}\ \bibnamefont {Souza}},
  \bibinfo {author} {\bibfnamefont {N.}~\bibnamefont {Rowley}}, \bibinfo
  {author} {\bibfnamefont {S.~J.}\ \bibnamefont {Sanders}}, \bibinfo {author}
  {\bibfnamefont {N.}~\bibnamefont {Aissaoui}}, \bibinfo {author}
  {\bibfnamefont {E.~E.}\ \bibnamefont {Alonso}}, \bibinfo {author}
  {\bibfnamefont {P.}~\bibnamefont {Bednarczyk}}, \bibinfo {author}
  {\bibfnamefont {N.}~\bibnamefont {Carlin}}, \bibinfo {author} {\bibfnamefont
  {S.}~\bibnamefont {Courtin}}, \bibinfo {author} {\bibfnamefont
  {A.}~\bibnamefont {Diaz-Torres}}, \bibinfo {author} {\bibfnamefont
  {A.}~\bibnamefont {Dummer}}, \bibinfo {author} {\bibfnamefont
  {F.}~\bibnamefont {Haas}}, \bibinfo {author} {\bibfnamefont {A.}~\bibnamefont
  {Hachem}}, \bibinfo {author} {\bibfnamefont {K.}~\bibnamefont {Hagino}},
  \bibinfo {author} {\bibfnamefont {F.}~\bibnamefont {Hoellinger}}, \bibinfo
  {author} {\bibfnamefont {R.~V.~F.}\ \bibnamefont {Janssens}}, \bibinfo
  {author} {\bibfnamefont {N.}~\bibnamefont {Kintz}}, \bibinfo {author}
  {\bibfnamefont {R.}~\bibnamefont {Liguori~Neto}}, \bibinfo {author}
  {\bibfnamefont {E.}~\bibnamefont {Martin}}, \bibinfo {author} {\bibfnamefont
  {M.~M.}\ \bibnamefont {Moura}}, \bibinfo {author} {\bibfnamefont {M.~G.}\
  \bibnamefont {Munhoz}}, \bibinfo {author} {\bibfnamefont {P.}~\bibnamefont
  {Papka}}, \bibinfo {author} {\bibfnamefont {M.}~\bibnamefont {Rousseau}},
  \bibinfo {author} {\bibfnamefont {A.}~\bibnamefont {S\`anchez~i Zafra}},
  \bibinfo {author} {\bibfnamefont {O.}~\bibnamefont {St\'ezowski}}, \bibinfo
  {author} {\bibfnamefont {A.~A.}\ \bibnamefont {Suaide}}, \bibinfo {author}
  {\bibfnamefont {E.~M.}\ \bibnamefont {Szanto}}, \bibinfo {author}
  {\bibfnamefont {A.}~\bibnamefont {Szanto~de Toledo}}, \bibinfo {author}
  {\bibfnamefont {S.}~\bibnamefont {Szilner}},\ and\ \bibinfo {author}
  {\bibfnamefont {J.}~\bibnamefont {Takahashi}},\ }\bibfield  {title} {\bibinfo
  {title} {Near-barrier fusion of weakly bound ${}^{6}\mathrm{Li}$ and
  ${}^{7}\mathrm{Li}$ nuclei with ${}^{59}\mathrm{Co}$},\ }\href
  {https://doi.org/10.1103/PhysRevC.67.054602} {\bibfield  {journal} {\bibinfo
  {journal} {Phys. Rev. C}\ }\textbf {\bibinfo {volume} {67}},\ \bibinfo
  {pages} {054602} (\bibinfo {year} {2003})}\BibitemShut {NoStop}%
\bibitem [{\citenamefont {Dasgupta}\ \emph {et~al.}(2004)\citenamefont
  {Dasgupta}, \citenamefont {Gomes}, \citenamefont {Hinde}, \citenamefont
  {Moraes}, \citenamefont {Anjos}, \citenamefont {Berriman}, \citenamefont
  {Butt}, \citenamefont {Carlin}, \citenamefont {Lubian}, \citenamefont
  {Morton}, \citenamefont {Newton},\ and\ \citenamefont {Szanto~de
  Toledo}}]{Dasgupta2004}%
  \BibitemOpen
  \bibfield  {author} {\bibinfo {author} {\bibfnamefont {M.}~\bibnamefont
  {Dasgupta}}, \bibinfo {author} {\bibfnamefont {P.~R.~S.}\ \bibnamefont
  {Gomes}}, \bibinfo {author} {\bibfnamefont {D.~J.}\ \bibnamefont {Hinde}},
  \bibinfo {author} {\bibfnamefont {S.~B.}\ \bibnamefont {Moraes}}, \bibinfo
  {author} {\bibfnamefont {R.~M.}\ \bibnamefont {Anjos}}, \bibinfo {author}
  {\bibfnamefont {A.~C.}\ \bibnamefont {Berriman}}, \bibinfo {author}
  {\bibfnamefont {R.~D.}\ \bibnamefont {Butt}}, \bibinfo {author}
  {\bibfnamefont {N.}~\bibnamefont {Carlin}}, \bibinfo {author} {\bibfnamefont
  {J.}~\bibnamefont {Lubian}}, \bibinfo {author} {\bibfnamefont {C.~R.}\
  \bibnamefont {Morton}}, \bibinfo {author} {\bibfnamefont {J.~O.}\
  \bibnamefont {Newton}},\ and\ \bibinfo {author} {\bibfnamefont
  {A.}~\bibnamefont {Szanto~de Toledo}},\ }\bibfield  {title} {\bibinfo {title}
  {Effect of breakup on the fusion of $^{6}\mathrm{Li}$, $^{7}\mathrm{Li}$, and
  $^{9}\mathrm{Be}$ with heavy nuclei},\ }\href
  {https://doi.org/10.1103/PhysRevC.70.024606} {\bibfield  {journal} {\bibinfo
  {journal} {Phys. Rev. C}\ }\textbf {\bibinfo {volume} {70}},\ \bibinfo
  {pages} {024606} (\bibinfo {year} {2004})}\BibitemShut {NoStop}%
\bibitem [{\citenamefont {Rath}\ \emph
  {et~al.}(2009{\natexlab{a}})\citenamefont {Rath}, \citenamefont {Santra},
  \citenamefont {Singh}, \citenamefont {Tripathi}, \citenamefont {Parkar},
  \citenamefont {Nayak}, \citenamefont {Mahata}, \citenamefont {Palit},
  \citenamefont {Kumar}, \citenamefont {Mukherjee}, \citenamefont
  {Appannababu},\ and\ \citenamefont {Choudhury}}]{Rath2009}%
  \BibitemOpen
  \bibfield  {author} {\bibinfo {author} {\bibfnamefont {P.~K.}\ \bibnamefont
  {Rath}}, \bibinfo {author} {\bibfnamefont {S.}~\bibnamefont {Santra}},
  \bibinfo {author} {\bibfnamefont {N.~L.}\ \bibnamefont {Singh}}, \bibinfo
  {author} {\bibfnamefont {R.}~\bibnamefont {Tripathi}}, \bibinfo {author}
  {\bibfnamefont {V.~V.}\ \bibnamefont {Parkar}}, \bibinfo {author}
  {\bibfnamefont {B.~K.}\ \bibnamefont {Nayak}}, \bibinfo {author}
  {\bibfnamefont {K.}~\bibnamefont {Mahata}}, \bibinfo {author} {\bibfnamefont
  {R.}~\bibnamefont {Palit}}, \bibinfo {author} {\bibfnamefont
  {S.}~\bibnamefont {Kumar}}, \bibinfo {author} {\bibfnamefont
  {S.}~\bibnamefont {Mukherjee}}, \bibinfo {author} {\bibfnamefont
  {S.}~\bibnamefont {Appannababu}},\ and\ \bibinfo {author} {\bibfnamefont
  {R.~K.}\ \bibnamefont {Choudhury}},\ }\bibfield  {title} {\bibinfo {title}
  {Suppression of complete fusion in the $^{6}\mathrm{Li}+^{144}\mathrm{Sm}$
  reaction},\ }\href {https://doi.org/10.1103/PhysRevC.79.051601} {\bibfield
  {journal} {\bibinfo  {journal} {Phys. Rev. C}\ }\textbf {\bibinfo {volume}
  {79}},\ \bibinfo {pages} {051601(R)} (\bibinfo {year}
  {2009}{\natexlab{a}})}\BibitemShut {NoStop}%
\bibitem [{\citenamefont {Gautam}\ \emph {et~al.}(2020)\citenamefont {Gautam},
  \citenamefont {Vinod}, \citenamefont {Duhan}, \citenamefont {Chahal},\ and\
  \citenamefont {Khatri}}]{Gautam2020121730}%
  \BibitemOpen
  \bibfield  {author} {\bibinfo {author} {\bibfnamefont {M.~S.}\ \bibnamefont
  {Gautam}}, \bibinfo {author} {\bibfnamefont {K.}~\bibnamefont {Vinod}},
  \bibinfo {author} {\bibfnamefont {S.}~\bibnamefont {Duhan}}, \bibinfo
  {author} {\bibfnamefont {R.~P.}\ \bibnamefont {Chahal}},\ and\ \bibinfo
  {author} {\bibfnamefont {H.}~\bibnamefont {Khatri}},\ }\bibfield  {title}
  {\bibinfo {title} {Investigation of contribution of complete and incomplete
  fusion in the total fusion process of $^{6}\mathrm{Li}$+$^{90,96}\mathrm{Zr}$
  reactions in vicinity of the coulomb barrier},\ }\href
  {https://doi.org/https://doi.org/10.1016/j.nuclphysa.2020.121730} {\bibfield
  {journal} {\bibinfo  {journal} {Nucl. Phys. A}\ }\textbf {\bibinfo {volume}
  {998}},\ \bibinfo {pages} {121730} (\bibinfo {year} {2020})}\BibitemShut
  {NoStop}%
\bibitem [{\citenamefont {Chetna}\ \emph {et~al.}(2022)\citenamefont {Chetna},
  \citenamefont {Shaikh}, \citenamefont {Singh},\ and\ \citenamefont
  {Kharab}}]{Chetna2022122418}%
  \BibitemOpen
  \bibfield  {author} {\bibinfo {author} {\bibnamefont {Chetna}}, \bibinfo
  {author} {\bibfnamefont {M.~M.}\ \bibnamefont {Shaikh}}, \bibinfo {author}
  {\bibfnamefont {P.}~\bibnamefont {Singh}},\ and\ \bibinfo {author}
  {\bibfnamefont {R.}~\bibnamefont {Kharab}},\ }\bibfield  {title} {\bibinfo
  {title} {Exploring the effects of breakup couplings for weakly bound
  projectile $^{9}\mathrm{Be}$ on various targets at around barrier energies},\
  }\href {https://doi.org/https://doi.org/10.1016/j.nuclphysa.2022.122418}
  {\bibfield  {journal} {\bibinfo  {journal} {Nucl. Phys. A}\ }\textbf
  {\bibinfo {volume} {1021}},\ \bibinfo {pages} {122418} (\bibinfo {year}
  {2022})}\BibitemShut {NoStop}%
\bibitem [{\citenamefont {Diaz-Torres}\ \emph {et~al.}(2003)\citenamefont
  {Diaz-Torres}, \citenamefont {Thompson},\ and\ \citenamefont
  {Beck}}]{Diaz-Torres2003}%
  \BibitemOpen
  \bibfield  {author} {\bibinfo {author} {\bibfnamefont {A.}~\bibnamefont
  {Diaz-Torres}}, \bibinfo {author} {\bibfnamefont {I.~J.}\ \bibnamefont
  {Thompson}},\ and\ \bibinfo {author} {\bibfnamefont {C.}~\bibnamefont
  {Beck}},\ }\bibfield  {title} {\bibinfo {title} {How does breakup influence
  the total fusion of $^{6,7}\mathrm{Li}$ at the coulomb barrier?},\ }\href
  {https://doi.org/10.1103/PhysRevC.68.044607} {\bibfield  {journal} {\bibinfo
  {journal} {Phys. Rev. C}\ }\textbf {\bibinfo {volume} {68}},\ \bibinfo
  {pages} {044607} (\bibinfo {year} {2003})}\BibitemShut {NoStop}%
\bibitem [{\citenamefont {Camacho}\ \emph {et~al.}(2019)\citenamefont
  {Camacho}, \citenamefont {Diaz-Torres},\ and\ \citenamefont
  {Zhang}}]{Camacho2019}%
  \BibitemOpen
  \bibfield  {author} {\bibinfo {author} {\bibfnamefont {A.~G.}\ \bibnamefont
  {Camacho}}, \bibinfo {author} {\bibfnamefont {A.}~\bibnamefont
  {Diaz-Torres}},\ and\ \bibinfo {author} {\bibfnamefont {H.~Q.}\ \bibnamefont
  {Zhang}},\ }\bibfield  {title} {\bibinfo {title} {Comparative study of the
  effect of resonances of the weakly bound nuclei $^{6,7}\mathrm{Li}$ on total
  fusion with light to heavy mass targets},\ }\href
  {https://doi.org/10.1103/PhysRevC.99.054615} {\bibfield  {journal} {\bibinfo
  {journal} {Phys. Rev. C}\ }\textbf {\bibinfo {volume} {99}},\ \bibinfo
  {pages} {054615} (\bibinfo {year} {2019})}\BibitemShut {NoStop}%
\bibitem [{\citenamefont {Lubian}\ \emph {et~al.}(2022)\citenamefont {Lubian},
  \citenamefont {Ferreira}, \citenamefont {Rangel}, \citenamefont {Cortes},\
  and\ \citenamefont {Canto}}]{Lubian2022}%
  \BibitemOpen
  \bibfield  {author} {\bibinfo {author} {\bibfnamefont {J.}~\bibnamefont
  {Lubian}}, \bibinfo {author} {\bibfnamefont {J.~L.}\ \bibnamefont
  {Ferreira}}, \bibinfo {author} {\bibfnamefont {J.}~\bibnamefont {Rangel}},
  \bibinfo {author} {\bibfnamefont {M.~R.}\ \bibnamefont {Cortes}},\ and\
  \bibinfo {author} {\bibfnamefont {L.~F.}\ \bibnamefont {Canto}},\ }\bibfield
  {title} {\bibinfo {title} {Fusion processes in collisions of
  $^{6}\mathrm{Li}$ beams on heavy targets},\ }\href
  {https://doi.org/10.1103/PhysRevC.105.054601} {\bibfield  {journal} {\bibinfo
   {journal} {Phys. Rev. C}\ }\textbf {\bibinfo {volume} {105}},\ \bibinfo
  {pages} {054601} (\bibinfo {year} {2022})}\BibitemShut {NoStop}%
\bibitem [{\citenamefont {de~Barbará}\ \emph {et~al.}(2007)\citenamefont
  {de~Barbará}, \citenamefont {Martí}, \citenamefont {Arazi}, \citenamefont
  {Capurro}, \citenamefont {Fernández~Niello}, \citenamefont {Figueira},
  \citenamefont {Pacheco}, \citenamefont {Ramírez}, \citenamefont
  {Rodríguez}, \citenamefont {Testoni}, \citenamefont {Verruno}, \citenamefont
  {Padrón}, \citenamefont {Gomes},\ and\ \citenamefont {Crema}}]{Barbara2007}%
  \BibitemOpen
  \bibfield  {author} {\bibinfo {author} {\bibfnamefont {E.}~\bibnamefont
  {de~Barbará}}, \bibinfo {author} {\bibfnamefont {G.~V.}\ \bibnamefont
  {Martí}}, \bibinfo {author} {\bibfnamefont {A.}~\bibnamefont {Arazi}},
  \bibinfo {author} {\bibfnamefont {O.~A.}\ \bibnamefont {Capurro}}, \bibinfo
  {author} {\bibfnamefont {J.~O.}\ \bibnamefont {Fernández~Niello}}, \bibinfo
  {author} {\bibfnamefont {J.~M.}\ \bibnamefont {Figueira}}, \bibinfo {author}
  {\bibfnamefont {A.~J.}\ \bibnamefont {Pacheco}}, \bibinfo {author}
  {\bibfnamefont {M.}~\bibnamefont {Ramírez}}, \bibinfo {author}
  {\bibfnamefont {M.~D.}\ \bibnamefont {Rodríguez}}, \bibinfo {author}
  {\bibfnamefont {J.~E.}\ \bibnamefont {Testoni}}, \bibinfo {author}
  {\bibfnamefont {M.}~\bibnamefont {Verruno}}, \bibinfo {author} {\bibfnamefont
  {I.}~\bibnamefont {Padrón}}, \bibinfo {author} {\bibfnamefont {P.~R.~S.}\
  \bibnamefont {Gomes}},\ and\ \bibinfo {author} {\bibfnamefont
  {E.}~\bibnamefont {Crema}},\ }\bibfield  {title} {\bibinfo {title} {Fusion
  cross sections for the 6,7li+27al, 9be+27al systems},\ }\href
  {https://doi.org/10.1063/1.2710578} {\bibfield  {journal} {\bibinfo
  {journal} {AIP Conference Proceedings}\ }\textbf {\bibinfo {volume} {884}},\
  \bibinfo {pages} {189} (\bibinfo {year} {2007})}\BibitemShut {NoStop}%
\bibitem [{\citenamefont {Sinha}\ and\ \citenamefont
  {Lubian}(2017)}]{Sinha2017a}%
  \BibitemOpen
  \bibfield  {author} {\bibinfo {author} {\bibfnamefont {M.}~\bibnamefont
  {Sinha}}\ and\ \bibinfo {author} {\bibfnamefont {J.}~\bibnamefont {Lubian}},\
  }\bibfield  {title} {\bibinfo {title} {Small suppression of the complete
  fusion of $^{6}\mathrm{Li}$ + $^{28}\mathrm{Si}$ reaction at near barrier
  energies},\ }\href {https://doi.org/10.1140/epja/i2017-12418-y} {\bibfield
  {journal} {\bibinfo  {journal} {Eur. Phys. J. A}\ }\textbf {\bibinfo {volume}
  {53}},\ \bibinfo {pages} {224} (\bibinfo {year} {2017})}\BibitemShut
  {NoStop}%
\bibitem [{\citenamefont {Palshetkar}\ \emph {et~al.}(2014)\citenamefont
  {Palshetkar}, \citenamefont {Thakur}, \citenamefont {Nanal}, \citenamefont
  {Shrivastava}, \citenamefont {Dokania}, \citenamefont {Singh}, \citenamefont
  {Parkar}, \citenamefont {Rout}, \citenamefont {Palit}, \citenamefont
  {Pillay}, \citenamefont {Bhattacharyya}, \citenamefont {Chatterjee},
  \citenamefont {Santra}, \citenamefont {Ramachandran},\ and\ \citenamefont
  {Singh}}]{Palshetkar2004}%
  \BibitemOpen
  \bibfield  {author} {\bibinfo {author} {\bibfnamefont {C.~S.}\ \bibnamefont
  {Palshetkar}}, \bibinfo {author} {\bibfnamefont {S.}~\bibnamefont {Thakur}},
  \bibinfo {author} {\bibfnamefont {V.}~\bibnamefont {Nanal}}, \bibinfo
  {author} {\bibfnamefont {A.}~\bibnamefont {Shrivastava}}, \bibinfo {author}
  {\bibfnamefont {N.}~\bibnamefont {Dokania}}, \bibinfo {author} {\bibfnamefont
  {V.}~\bibnamefont {Singh}}, \bibinfo {author} {\bibfnamefont {V.~V.}\
  \bibnamefont {Parkar}}, \bibinfo {author} {\bibfnamefont {P.~C.}\
  \bibnamefont {Rout}}, \bibinfo {author} {\bibfnamefont {R.}~\bibnamefont
  {Palit}}, \bibinfo {author} {\bibfnamefont {R.~G.}\ \bibnamefont {Pillay}},
  \bibinfo {author} {\bibfnamefont {S.}~\bibnamefont {Bhattacharyya}}, \bibinfo
  {author} {\bibfnamefont {A.}~\bibnamefont {Chatterjee}}, \bibinfo {author}
  {\bibfnamefont {S.}~\bibnamefont {Santra}}, \bibinfo {author} {\bibfnamefont
  {K.}~\bibnamefont {Ramachandran}},\ and\ \bibinfo {author} {\bibfnamefont
  {N.~L.}\ \bibnamefont {Singh}},\ }\bibfield  {title} {\bibinfo {title}
  {Fusion and quasi-elastic scattering in the
  $^{6,7}\mathrm{Li}$+$^{197}\mathrm{Au}$ systems},\ }\href
  {https://doi.org/10.1103/PhysRevC.89.024607} {\bibfield  {journal} {\bibinfo
  {journal} {Phys. Rev. C}\ }\textbf {\bibinfo {volume} {89}},\ \bibinfo
  {pages} {024607} (\bibinfo {year} {2014})}\BibitemShut {NoStop}%
\bibitem [{\citenamefont {Dasgupta}\ \emph {et~al.}(2002)\citenamefont
  {Dasgupta}, \citenamefont {Hinde}, \citenamefont {Hagino}, \citenamefont
  {Moraes}, \citenamefont {Gomes}, \citenamefont {Anjos}, \citenamefont {Butt},
  \citenamefont {Berriman}, \citenamefont {Carlin}, \citenamefont {Morton},
  \citenamefont {Newton},\ and\ \citenamefont {Szanto~de
  Toledo}}]{Dasgupta2002}%
  \BibitemOpen
  \bibfield  {author} {\bibinfo {author} {\bibfnamefont {M.}~\bibnamefont
  {Dasgupta}}, \bibinfo {author} {\bibfnamefont {D.~J.}\ \bibnamefont {Hinde}},
  \bibinfo {author} {\bibfnamefont {K.}~\bibnamefont {Hagino}}, \bibinfo
  {author} {\bibfnamefont {S.~B.}\ \bibnamefont {Moraes}}, \bibinfo {author}
  {\bibfnamefont {P.~R.~S.}\ \bibnamefont {Gomes}}, \bibinfo {author}
  {\bibfnamefont {R.~M.}\ \bibnamefont {Anjos}}, \bibinfo {author}
  {\bibfnamefont {R.~D.}\ \bibnamefont {Butt}}, \bibinfo {author}
  {\bibfnamefont {A.~C.}\ \bibnamefont {Berriman}}, \bibinfo {author}
  {\bibfnamefont {N.}~\bibnamefont {Carlin}}, \bibinfo {author} {\bibfnamefont
  {C.~R.}\ \bibnamefont {Morton}}, \bibinfo {author} {\bibfnamefont {J.~O.}\
  \bibnamefont {Newton}},\ and\ \bibinfo {author} {\bibfnamefont
  {A.}~\bibnamefont {Szanto~de Toledo}},\ }\bibfield  {title} {\bibinfo {title}
  {Fusion and breakup in the reactions of ${}^{6}\mathrm{Li}$ and
  ${}^{7}\mathrm{Li}$ nuclei with ${}^{209}\mathrm{Bi}$},\ }\href
  {https://doi.org/10.1103/PhysRevC.66.041602} {\bibfield  {journal} {\bibinfo
  {journal} {Phys. Rev. C}\ }\textbf {\bibinfo {volume} {66}},\ \bibinfo
  {pages} {041602(R)} (\bibinfo {year} {2002})}\BibitemShut {NoStop}%
\bibitem [{\citenamefont {Hagino}\ \emph {et~al.}(2000)\citenamefont {Hagino},
  \citenamefont {Vitturi}, \citenamefont {Dasso},\ and\ \citenamefont
  {Lenzi}}]{Hagino2000}%
  \BibitemOpen
  \bibfield  {author} {\bibinfo {author} {\bibfnamefont {K.}~\bibnamefont
  {Hagino}}, \bibinfo {author} {\bibfnamefont {A.}~\bibnamefont {Vitturi}},
  \bibinfo {author} {\bibfnamefont {C.~H.}\ \bibnamefont {Dasso}},\ and\
  \bibinfo {author} {\bibfnamefont {S.~M.}\ \bibnamefont {Lenzi}},\ }\bibfield
  {title} {\bibinfo {title} {Role of breakup processes in fusion enhancement of
  drip-line nuclei at energies below the coulomb barrier},\ }\href
  {https://doi.org/10.1103/PhysRevC.61.037602} {\bibfield  {journal} {\bibinfo
  {journal} {Phys. Rev. C}\ }\textbf {\bibinfo {volume} {61}},\ \bibinfo
  {pages} {037602} (\bibinfo {year} {2000})}\BibitemShut {NoStop}%
\bibitem [{\citenamefont {Diaz-Torres}\ and\ \citenamefont
  {Thompson}(2002)}]{Diaz-Torres2002}%
  \BibitemOpen
  \bibfield  {author} {\bibinfo {author} {\bibfnamefont {A.}~\bibnamefont
  {Diaz-Torres}}\ and\ \bibinfo {author} {\bibfnamefont {I.~J.}\ \bibnamefont
  {Thompson}},\ }\bibfield  {title} {\bibinfo {title} {Effect of continuum
  couplings in fusion of halo ${}^{11}\mathrm{Be}$ on ${}^{208}\mathrm{Pb}$
  around the coulomb barrier},\ }\href
  {https://doi.org/10.1103/PhysRevC.65.024606} {\bibfield  {journal} {\bibinfo
  {journal} {Phys. Rev. C}\ }\textbf {\bibinfo {volume} {65}},\ \bibinfo
  {pages} {024606} (\bibinfo {year} {2002})}\BibitemShut {NoStop}%
\bibitem [{\citenamefont {Hashimoto}\ \emph {et~al.}(2009)\citenamefont
  {Hashimoto}, \citenamefont {Ogata}, \citenamefont {Chiba},\ and\
  \citenamefont {Yahiro}}]{Hashimoto2009}%
  \BibitemOpen
  \bibfield  {author} {\bibinfo {author} {\bibfnamefont {S.}~\bibnamefont
  {Hashimoto}}, \bibinfo {author} {\bibfnamefont {K.}~\bibnamefont {Ogata}},
  \bibinfo {author} {\bibfnamefont {S.}~\bibnamefont {Chiba}},\ and\ \bibinfo
  {author} {\bibfnamefont {M.}~\bibnamefont {Yahiro}},\ }\bibfield  {title}
  {\bibinfo {title} {{New Approach for Evaluating Incomplete and Complete
  Fusion Cross Sections with Continuum-Discretized Coupled-Channels Method}},\
  }\href {https://doi.org/10.1143/PTP.122.1291} {\bibfield  {journal} {\bibinfo
   {journal} {Prog. Theor. Phys.}\ }\textbf {\bibinfo {volume} {122}},\
  \bibinfo {pages} {1291} (\bibinfo {year} {2009})}\BibitemShut {NoStop}%
\bibitem [{\citenamefont {Parkar}\ \emph {et~al.}(2016)\citenamefont {Parkar},
  \citenamefont {Jha},\ and\ \citenamefont {Kailas}}]{Parkar2016}%
  \BibitemOpen
  \bibfield  {author} {\bibinfo {author} {\bibfnamefont {V.~V.}\ \bibnamefont
  {Parkar}}, \bibinfo {author} {\bibfnamefont {V.}~\bibnamefont {Jha}},\ and\
  \bibinfo {author} {\bibfnamefont {S.}~\bibnamefont {Kailas}},\ }\bibfield
  {title} {\bibinfo {title} {Exploring contributions from incomplete fusion in
  $^{6,7}\mathrm{Li}+^{209}\mathrm{Bi}$ and
  $^{6,7}\mathrm{Li}+^{198}\mathrm{Pt}$ reactions},\ }\href
  {https://doi.org/10.1103/PhysRevC.94.024609} {\bibfield  {journal} {\bibinfo
  {journal} {Phys. Rev. C}\ }\textbf {\bibinfo {volume} {94}},\ \bibinfo
  {pages} {024609} (\bibinfo {year} {2016})}\BibitemShut {NoStop}%
\bibitem [{\citenamefont {G\'omez~Camacho}\ \emph {et~al.}(2018)\citenamefont
  {G\'omez~Camacho}, \citenamefont {Wang},\ and\ \citenamefont
  {Zhang}}]{Gomez2018}%
  \BibitemOpen
  \bibfield  {author} {\bibinfo {author} {\bibfnamefont {A.}~\bibnamefont
  {G\'omez~Camacho}}, \bibinfo {author} {\bibfnamefont {B.}~\bibnamefont
  {Wang}},\ and\ \bibinfo {author} {\bibfnamefont {H.~Q.}\ \bibnamefont
  {Zhang}},\ }\bibfield  {title} {\bibinfo {title} {Systematic
  continuum-discretized coupled-channels calculations of total fusion for
  $^{6}\mathrm{Li}$ with targets $^{28}\mathrm{Si}, ^{59}\mathrm{Co},
  ^{96}\mathrm{Zr}, ^{198}\mathrm{Pt}$, and $^{209}\mathrm{Bi}$: Effect of
  resonance states},\ }\href {https://doi.org/10.1103/PhysRevC.97.054610}
  {\bibfield  {journal} {\bibinfo  {journal} {Phys. Rev. C}\ }\textbf {\bibinfo
  {volume} {97}},\ \bibinfo {pages} {054610} (\bibinfo {year}
  {2018})}\BibitemShut {NoStop}%
\bibitem [{\citenamefont {Rangel}\ \emph {et~al.}(2020)\citenamefont {Rangel},
  \citenamefont {Cortes}, \citenamefont {Lubian},\ and\ \citenamefont
  {Canto}}]{Rangel2020}%
  \BibitemOpen
  \bibfield  {author} {\bibinfo {author} {\bibfnamefont {J.}~\bibnamefont
  {Rangel}}, \bibinfo {author} {\bibfnamefont {M.}~\bibnamefont {Cortes}},
  \bibinfo {author} {\bibfnamefont {J.}~\bibnamefont {Lubian}},\ and\ \bibinfo
  {author} {\bibfnamefont {L.}~\bibnamefont {Canto}},\ }\bibfield  {title}
  {\bibinfo {title} {Theory of complete and incomplete fusion of weakly bound
  systems},\ }\href
  {https://doi.org/https://doi.org/10.1016/j.physletb.2020.135337} {\bibfield
  {journal} {\bibinfo  {journal} {Physics Letters B}\ }\textbf {\bibinfo
  {volume} {803}},\ \bibinfo {pages} {135337} (\bibinfo {year}
  {2020})}\BibitemShut {NoStop}%
\bibitem [{\citenamefont {Cortes}\ \emph {et~al.}(2020)\citenamefont {Cortes},
  \citenamefont {Rangel}, \citenamefont {Ferreira}, \citenamefont {Lubian},\
  and\ \citenamefont {Canto}}]{Cortes2020}%
  \BibitemOpen
  \bibfield  {author} {\bibinfo {author} {\bibfnamefont {M.~R.}\ \bibnamefont
  {Cortes}}, \bibinfo {author} {\bibfnamefont {J.}~\bibnamefont {Rangel}},
  \bibinfo {author} {\bibfnamefont {J.~L.}\ \bibnamefont {Ferreira}}, \bibinfo
  {author} {\bibfnamefont {J.}~\bibnamefont {Lubian}},\ and\ \bibinfo {author}
  {\bibfnamefont {L.~F.}\ \bibnamefont {Canto}},\ }\bibfield  {title} {\bibinfo
  {title} {Complete and incomplete fusion of $^{7}\mathrm{Li}$ projectiles on
  heavy targets},\ }\href {https://doi.org/10.1103/PhysRevC.102.064628}
  {\bibfield  {journal} {\bibinfo  {journal} {Phys. Rev. C}\ }\textbf {\bibinfo
  {volume} {102}},\ \bibinfo {pages} {064628} (\bibinfo {year}
  {2020})}\BibitemShut {NoStop}%
\bibitem [{\citenamefont {Yahiro}\ \emph {et~al.}(1986)\citenamefont {Yahiro},
  \citenamefont {Iseri}, \citenamefont {Kameyama}, \citenamefont {Kamimura},\
  and\ \citenamefont {Kawai}}]{Yahiro1986}%
  \BibitemOpen
  \bibfield  {author} {\bibinfo {author} {\bibfnamefont {M.}~\bibnamefont
  {Yahiro}}, \bibinfo {author} {\bibfnamefont {Y.}~\bibnamefont {Iseri}},
  \bibinfo {author} {\bibfnamefont {H.}~\bibnamefont {Kameyama}}, \bibinfo
  {author} {\bibfnamefont {M.}~\bibnamefont {Kamimura}},\ and\ \bibinfo
  {author} {\bibfnamefont {M.}~\bibnamefont {Kawai}},\ }\bibfield  {title}
  {\bibinfo {title} {{Chapter III. Effects of Deuteron Virtual Breakup on
  Deuteron Elastic and Inelastic Scattering}},\ }\href
  {https://doi.org/10.1143/PTPS.89.32} {\bibfield  {journal} {\bibinfo
  {journal} {Prog. Theor. Phys. Suppl.}\ }\textbf {\bibinfo {volume} {89}},\
  \bibinfo {pages} {32} (\bibinfo {year} {1986})}\BibitemShut {NoStop}%
\bibitem [{\citenamefont {Ogata}\ and\ \citenamefont
  {Yoshida}(2016)}]{Ogata2016}%
  \BibitemOpen
  \bibfield  {author} {\bibinfo {author} {\bibfnamefont {K.}~\bibnamefont
  {Ogata}}\ and\ \bibinfo {author} {\bibfnamefont {K.}~\bibnamefont
  {Yoshida}},\ }\bibfield  {title} {\bibinfo {title} {Applicability of the
  continuum-discretized coupled-channels method to the deuteron breakup at low
  energies},\ }\href {https://doi.org/10.1103/PhysRevC.94.051603} {\bibfield
  {journal} {\bibinfo  {journal} {Phys. Rev. C}\ }\textbf {\bibinfo {volume}
  {94}},\ \bibinfo {pages} {051603(R)} (\bibinfo {year} {2016})}\BibitemShut
  {NoStop}%
\bibitem [{\citenamefont {Chen}\ \emph {et~al.}(2022)\citenamefont {Chen},
  \citenamefont {Guo}, \citenamefont {Ye}, \citenamefont {Ying}, \citenamefont
  {Sun},\ and\ \citenamefont {Han}}]{Chen2022}%
  \BibitemOpen
  \bibfield  {author} {\bibinfo {author} {\bibfnamefont {W.}~\bibnamefont
  {Chen}}, \bibinfo {author} {\bibfnamefont {H.}~\bibnamefont {Guo}}, \bibinfo
  {author} {\bibfnamefont {T.}~\bibnamefont {Ye}}, \bibinfo {author}
  {\bibfnamefont {Y.}~\bibnamefont {Ying}}, \bibinfo {author} {\bibfnamefont
  {W.}~\bibnamefont {Sun}},\ and\ \bibinfo {author} {\bibfnamefont
  {Y.}~\bibnamefont {Han}},\ }\bibfield  {title} {\bibinfo {title} {Application
  of the lagrange-mesh method in continuum-discretized coupled-channel
  calculations},\ }\href {https://doi.org/10.1088/1361-6471/ac7249} {\bibfield
  {journal} {\bibinfo  {journal} {J. Phys. G: Nucl. Part. Phys.}\ }\textbf
  {\bibinfo {volume} {49}},\ \bibinfo {pages} {075104} (\bibinfo {year}
  {2022})}\BibitemShut {NoStop}%
\bibitem [{\citenamefont {Wilczyński}(1973)}]{Wilczynski1973386}%
  \BibitemOpen
  \bibfield  {author} {\bibinfo {author} {\bibfnamefont {J.}~\bibnamefont
  {Wilczyński}},\ }\bibfield  {title} {\bibinfo {title} {Calculations of the
  critical angular momentum in the entrance reaction channel},\ }\href
  {https://doi.org/https://doi.org/10.1016/0375-9474(73)90474-0} {\bibfield
  {journal} {\bibinfo  {journal} {Nucl. Phys. A}\ }\textbf {\bibinfo {volume}
  {216}},\ \bibinfo {pages} {386} (\bibinfo {year} {1973})}\BibitemShut
  {NoStop}%
\bibitem [{\citenamefont {Wilczy\ifmmode~\acute{n}\else \'{n}\fi{}ski}\ \emph
  {et~al.}(1980)\citenamefont {Wilczy\ifmmode~\acute{n}\else \'{n}\fi{}ski},
  \citenamefont {Siwek-Wilczy\ifmmode~\acute{n}\else \'{n}\fi{}ska},
  \citenamefont {van Driel}, \citenamefont {Gonggrijp}, \citenamefont
  {Hageman}, \citenamefont {Janssens}, \citenamefont {\L{}ukasiak},\ and\
  \citenamefont {Siemssen}}]{Wilczynski1980}%
  \BibitemOpen
  \bibfield  {author} {\bibinfo {author} {\bibfnamefont {J.}~\bibnamefont
  {Wilczy\ifmmode~\acute{n}\else \'{n}\fi{}ski}}, \bibinfo {author}
  {\bibfnamefont {K.}~\bibnamefont {Siwek-Wilczy\ifmmode~\acute{n}\else
  \'{n}\fi{}ska}}, \bibinfo {author} {\bibfnamefont {J.}~\bibnamefont {van
  Driel}}, \bibinfo {author} {\bibfnamefont {S.}~\bibnamefont {Gonggrijp}},
  \bibinfo {author} {\bibfnamefont {D.~C. J.~M.}\ \bibnamefont {Hageman}},
  \bibinfo {author} {\bibfnamefont {R.~V.~F.}\ \bibnamefont {Janssens}},
  \bibinfo {author} {\bibfnamefont {J.}~\bibnamefont {\L{}ukasiak}},\ and\
  \bibinfo {author} {\bibfnamefont {R.~H.}\ \bibnamefont {Siemssen}},\
  }\bibfield  {title} {\bibinfo {title} {Incomplete-fusion reactions in the
  $^{14}\mathrm{N}$+$^{159}\mathrm{Tb}$ system and a "sum-rule model" for
  fusion and incomplete-fusion reactions},\ }\href
  {https://doi.org/10.1103/PhysRevLett.45.606} {\bibfield  {journal} {\bibinfo
  {journal} {Phys. Rev. Lett.}\ }\textbf {\bibinfo {volume} {45}},\ \bibinfo
  {pages} {606} (\bibinfo {year} {1980})}\BibitemShut {NoStop}%
\bibitem [{\citenamefont {Wilczynski}\ \emph {et~al.}(1982)\citenamefont
  {Wilczynski}, \citenamefont {Siwek-Wilczynska}, \citenamefont {{Van Driel}},
  \citenamefont {Gonggrijp}, \citenamefont {Hageman}, \citenamefont {Janssens},
  \citenamefont {Łukasiak}, \citenamefont {Siemssen},\ and\ \citenamefont
  {{Van Der Werf}}}]{Wilczynski1982109}%
  \BibitemOpen
  \bibfield  {author} {\bibinfo {author} {\bibfnamefont {J.}~\bibnamefont
  {Wilczynski}}, \bibinfo {author} {\bibfnamefont {K.}~\bibnamefont
  {Siwek-Wilczynska}}, \bibinfo {author} {\bibfnamefont {J.}~\bibnamefont {{Van
  Driel}}}, \bibinfo {author} {\bibfnamefont {S.}~\bibnamefont {Gonggrijp}},
  \bibinfo {author} {\bibfnamefont {D.}~\bibnamefont {Hageman}}, \bibinfo
  {author} {\bibfnamefont {R.}~\bibnamefont {Janssens}}, \bibinfo {author}
  {\bibfnamefont {J.}~\bibnamefont {Łukasiak}}, \bibinfo {author}
  {\bibfnamefont {R.}~\bibnamefont {Siemssen}},\ and\ \bibinfo {author}
  {\bibfnamefont {S.}~\bibnamefont {{Van Der Werf}}},\ }\bibfield  {title}
  {\bibinfo {title} {Binary l-matched reactions in $^{14}\mathrm{N}$ +
  $^{159}\mathrm{Tb}$ collisions},\ }\href
  {https://doi.org/https://doi.org/10.1016/0375-9474(82)90183-X} {\bibfield
  {journal} {\bibinfo  {journal} {Nucl. Phys. A}\ }\textbf {\bibinfo {volume}
  {373}},\ \bibinfo {pages} {109} (\bibinfo {year} {1982})}\BibitemShut
  {NoStop}%
\bibitem [{\citenamefont {Mukeru}\ \emph {et~al.}(2020)\citenamefont {Mukeru},
  \citenamefont {Lekala}, \citenamefont {Lubian},\ and\ \citenamefont
  {Tomio}}]{Mukeru2020121700}%
  \BibitemOpen
  \bibfield  {author} {\bibinfo {author} {\bibfnamefont {B.}~\bibnamefont
  {Mukeru}}, \bibinfo {author} {\bibfnamefont {M.}~\bibnamefont {Lekala}},
  \bibinfo {author} {\bibfnamefont {J.}~\bibnamefont {Lubian}},\ and\ \bibinfo
  {author} {\bibfnamefont {L.}~\bibnamefont {Tomio}},\ }\bibfield  {title}
  {\bibinfo {title} {Theoretical analysis of $^{8}\mathrm{Li}$ +
  $^{208}\mathrm{Pb}$ reaction and the critical angular momentum for complete
  fusion},\ }\href
  {https://doi.org/https://doi.org/10.1016/j.nuclphysa.2020.121700} {\bibfield
  {journal} {\bibinfo  {journal} {Nucl. Phys. A}\ }\textbf {\bibinfo {volume}
  {996}},\ \bibinfo {pages} {121700} (\bibinfo {year} {2020})}\BibitemShut
  {NoStop}%
\bibitem [{\citenamefont {Mukeru}\ and\ \citenamefont
  {Tomio}(2021)}]{Mukeru2021}%
  \BibitemOpen
  \bibfield  {author} {\bibinfo {author} {\bibfnamefont {B.}~\bibnamefont
  {Mukeru}}\ and\ \bibinfo {author} {\bibfnamefont {L.}~\bibnamefont {Tomio}},\
  }\bibfield  {title} {\bibinfo {title} {Analysis of fusion cross sections in
  the $^{9}\mathrm{Be}$ projectile breakup on different target nuclei},\ }\href
  {https://doi.org/10.1007/s13538-020-00797-0} {\bibfield  {journal} {\bibinfo
  {journal} {Braz J Phys}\ }\textbf {\bibinfo {volume} {51}},\ \bibinfo {pages}
  {157} (\bibinfo {year} {2021})}\BibitemShut {NoStop}%
\bibitem [{\citenamefont {Tilley}\ \emph {et~al.}(2002)\citenamefont {Tilley},
  \citenamefont {Cheves}, \citenamefont {Godwin}, \citenamefont {Hale},
  \citenamefont {Hofmann}, \citenamefont {Kelley}, \citenamefont {Sheu},\ and\
  \citenamefont {Weller}}]{Tilley20023}%
  \BibitemOpen
  \bibfield  {author} {\bibinfo {author} {\bibfnamefont {D.}~\bibnamefont
  {Tilley}}, \bibinfo {author} {\bibfnamefont {C.}~\bibnamefont {Cheves}},
  \bibinfo {author} {\bibfnamefont {J.}~\bibnamefont {Godwin}}, \bibinfo
  {author} {\bibfnamefont {G.}~\bibnamefont {Hale}}, \bibinfo {author}
  {\bibfnamefont {H.}~\bibnamefont {Hofmann}}, \bibinfo {author} {\bibfnamefont
  {J.}~\bibnamefont {Kelley}}, \bibinfo {author} {\bibfnamefont
  {C.}~\bibnamefont {Sheu}},\ and\ \bibinfo {author} {\bibfnamefont
  {H.}~\bibnamefont {Weller}},\ }\bibfield  {title} {\bibinfo {title} {{Energy
  levels of light nuclei A=5, 6, 7}},\ }\href
  {https://doi.org/https://doi.org/10.1016/S0375-9474(02)00597-3} {\bibfield
  {journal} {\bibinfo  {journal} {Nucl. Phys. A}\ }\textbf {\bibinfo {volume}
  {708}},\ \bibinfo {pages} {3} (\bibinfo {year} {2002})}\BibitemShut {NoStop}%
\bibitem [{\citenamefont {Schmelzbach}\ \emph {et~al.}(1972)\citenamefont
  {Schmelzbach}, \citenamefont {Gr{\"{u}}ebler}, \citenamefont {K{\"{o}}nig},\
  and\ \citenamefont {Marmier}}]{Schmelzbach1972}%
  \BibitemOpen
  \bibfield  {author} {\bibinfo {author} {\bibfnamefont {P.}~\bibnamefont
  {Schmelzbach}}, \bibinfo {author} {\bibfnamefont {W.}~\bibnamefont
  {Gr{\"{u}}ebler}}, \bibinfo {author} {\bibfnamefont {V.}~\bibnamefont
  {K{\"{o}}nig}},\ and\ \bibinfo {author} {\bibfnamefont {P.}~\bibnamefont
  {Marmier}},\ }\bibfield  {title} {\bibinfo {title} {Phase-shift analysis of
  d-$\alpha$ elastic scattering},\ }\href
  {https://doi.org/https://doi.org/10.1016/0375-9474(72)90453-8} {\bibfield
  {journal} {\bibinfo  {journal} {Nucl. Phys. A}\ }\textbf {\bibinfo {volume}
  {184}},\ \bibinfo {pages} {193} (\bibinfo {year} {1972})}\BibitemShut
  {NoStop}%
\bibitem [{\citenamefont {Gr{\"{u}}ebler}\ \emph {et~al.}(1975)\citenamefont
  {Gr{\"{u}}ebler}, \citenamefont {Schmelzbach}, \citenamefont {K{\"{o}}nig},
  \citenamefont {Risler},\ and\ \citenamefont {Boerma}}]{Gruebler1975}%
  \BibitemOpen
  \bibfield  {author} {\bibinfo {author} {\bibfnamefont {W.}~\bibnamefont
  {Gr{\"{u}}ebler}}, \bibinfo {author} {\bibfnamefont {P.}~\bibnamefont
  {Schmelzbach}}, \bibinfo {author} {\bibfnamefont {V.}~\bibnamefont
  {K{\"{o}}nig}}, \bibinfo {author} {\bibfnamefont {R.}~\bibnamefont
  {Risler}},\ and\ \bibinfo {author} {\bibfnamefont {D.}~\bibnamefont
  {Boerma}},\ }\bibfield  {title} {\bibinfo {title} {Phase-shift analysis of
  d-$\alpha$ elastic scattering between 3 and 17 $\mathrm{MeV}$},\ }\href
  {https://doi.org/https://doi.org/10.1016/0375-9474(75)90048-2} {\bibfield
  {journal} {\bibinfo  {journal} {Nucl. Phys. A}\ }\textbf {\bibinfo {volume}
  {242}},\ \bibinfo {pages} {265} (\bibinfo {year} {1975})}\BibitemShut
  {NoStop}%
\bibitem [{\citenamefont {Jenny}\ \emph {et~al.}(1983)\citenamefont {Jenny},
  \citenamefont {Gr{\"{u}}ebler}, \citenamefont {K{\"{o}}nig}, \citenamefont
  {Schmelzbach},\ and\ \citenamefont {Schweizer}}]{Jenny1983}%
  \BibitemOpen
  \bibfield  {author} {\bibinfo {author} {\bibfnamefont {B.}~\bibnamefont
  {Jenny}}, \bibinfo {author} {\bibfnamefont {W.}~\bibnamefont
  {Gr{\"{u}}ebler}}, \bibinfo {author} {\bibfnamefont {V.}~\bibnamefont
  {K{\"{o}}nig}}, \bibinfo {author} {\bibfnamefont {P.}~\bibnamefont
  {Schmelzbach}},\ and\ \bibinfo {author} {\bibfnamefont {C.}~\bibnamefont
  {Schweizer}},\ }\bibfield  {title} {\bibinfo {title} {Phase-shift analysis of
  d-$\alpha$ elastic scattering between 3 and 43 $\mathrm{MeV}$},\ }\href
  {https://doi.org/https://doi.org/10.1016/0375-9474(83)90078-7} {\bibfield
  {journal} {\bibinfo  {journal} {Nucl. Phys. A}\ }\textbf {\bibinfo {volume}
  {397}},\ \bibinfo {pages} {61} (\bibinfo {year} {1983})}\BibitemShut
  {NoStop}%
\bibitem [{\citenamefont {Saito}(1969)}]{Saito1969}%
  \BibitemOpen
  \bibfield  {author} {\bibinfo {author} {\bibfnamefont {S.}~\bibnamefont
  {Saito}},\ }\bibfield  {title} {\bibinfo {title} {{Interaction between
  Clusters and Pauli Principle*)}},\ }\href
  {https://doi.org/10.1143/PTP.41.705} {\bibfield  {journal} {\bibinfo
  {journal} {Progress of Theoretical Physics}\ }\textbf {\bibinfo {volume}
  {41}},\ \bibinfo {pages} {705} (\bibinfo {year} {1969})}\BibitemShut
  {NoStop}%
\bibitem [{\citenamefont {Horiuchi}(1977)}]{Horiuchi1977}%
  \BibitemOpen
  \bibfield  {author} {\bibinfo {author} {\bibfnamefont {H.}~\bibnamefont
  {Horiuchi}},\ }\bibfield  {title} {\bibinfo {title} {{Chapter III. Kernels of
  GCM, RGM and OCM and Their Calculation Methods}},\ }\href
  {https://doi.org/10.1143/PTPS.62.90} {\bibfield  {journal} {\bibinfo
  {journal} {Progress of Theoretical Physics Supplement}\ }\textbf {\bibinfo
  {volume} {62}},\ \bibinfo {pages} {90} (\bibinfo {year} {1977})}\BibitemShut
  {NoStop}%
\bibitem [{\citenamefont {Sakuragi}\ \emph {et~al.}(1986)\citenamefont
  {Sakuragi}, \citenamefont {Yahiro},\ and\ \citenamefont
  {Kamimura}}]{Sakuragi1986}%
  \BibitemOpen
  \bibfield  {author} {\bibinfo {author} {\bibfnamefont {Y.}~\bibnamefont
  {Sakuragi}}, \bibinfo {author} {\bibfnamefont {M.}~\bibnamefont {Yahiro}},\
  and\ \bibinfo {author} {\bibfnamefont {M.}~\bibnamefont {Kamimura}},\
  }\bibfield  {title} {\bibinfo {title} {{Chapter VI. Microscopic
  Coupled-Channels Study of Scattering and Breakup of Light Heavy-Ions}},\
  }\href {https://doi.org/10.1143/PTPS.89.136} {\bibfield  {journal} {\bibinfo
  {journal} {Prog. Theor. Phys. Suppl.}\ }\textbf {\bibinfo {volume} {89}},\
  \bibinfo {pages} {136} (\bibinfo {year} {1986})}\BibitemShut {NoStop}%
\bibitem [{\citenamefont {Guo}\ \emph {et~al.}(2013)\citenamefont {Guo},
  \citenamefont {Watanabe}, \citenamefont {Matsumoto}, \citenamefont {Ogata},\
  and\ \citenamefont {Yahiro}}]{Guo2014}%
  \BibitemOpen
  \bibfield  {author} {\bibinfo {author} {\bibfnamefont {H.}~\bibnamefont
  {Guo}}, \bibinfo {author} {\bibfnamefont {Y.}~\bibnamefont {Watanabe}},
  \bibinfo {author} {\bibfnamefont {T.}~\bibnamefont {Matsumoto}}, \bibinfo
  {author} {\bibfnamefont {K.}~\bibnamefont {Ogata}},\ and\ \bibinfo {author}
  {\bibfnamefont {M.}~\bibnamefont {Yahiro}},\ }\bibfield  {title} {\bibinfo
  {title} {Systematic analysis of nucleon scattering from $^{6,7}\mathrm{Li}$
  with the continuum discretized coupled channels method},\ }\href
  {https://doi.org/10.1103/PhysRevC.87.024610} {\bibfield  {journal} {\bibinfo
  {journal} {Phys. Rev. C}\ }\textbf {\bibinfo {volume} {87}},\ \bibinfo
  {pages} {024610} (\bibinfo {year} {2013})}\BibitemShut {NoStop}%
\bibitem [{\citenamefont {Koning}\ and\ \citenamefont
  {Delaroche}(2003)}]{Koning2003231}%
  \BibitemOpen
  \bibfield  {author} {\bibinfo {author} {\bibfnamefont {A.}~\bibnamefont
  {Koning}}\ and\ \bibinfo {author} {\bibfnamefont {J.}~\bibnamefont
  {Delaroche}},\ }\bibfield  {title} {\bibinfo {title} {Local and global
  nucleon optical models from 1 $\mathrm{KeV}$ to 200 $\mathrm{MeV}$},\ }\href
  {https://doi.org/https://doi.org/10.1016/S0375-9474(02)01321-0} {\bibfield
  {journal} {\bibinfo  {journal} {Nucl. Phys. A}\ }\textbf {\bibinfo {volume}
  {713}},\ \bibinfo {pages} {231} (\bibinfo {year} {2003})}\BibitemShut
  {NoStop}%
\bibitem [{\citenamefont {Bergstrom}(1979)}]{Bergstrom1979458}%
  \BibitemOpen
  \bibfield  {author} {\bibinfo {author} {\bibfnamefont {J.}~\bibnamefont
  {Bergstrom}},\ }\bibfield  {title} {\bibinfo {title} {6li electromagnetic
  form factors and phenomenological cluster models},\ }\href
  {https://doi.org/https://doi.org/10.1016/0375-9474(79)90269-0} {\bibfield
  {journal} {\bibinfo  {journal} {Nuclear Physics A}\ }\textbf {\bibinfo
  {volume} {327}},\ \bibinfo {pages} {458} (\bibinfo {year}
  {1979})}\BibitemShut {NoStop}%
\bibitem [{\citenamefont {Tanihata}\ \emph {et~al.}(1985)\citenamefont
  {Tanihata}, \citenamefont {Hamagaki}, \citenamefont {Hashimoto},
  \citenamefont {Shida}, \citenamefont {Yoshikawa}, \citenamefont {Sugimoto},
  \citenamefont {Yamakawa}, \citenamefont {Kobayashi},\ and\ \citenamefont
  {Takahashi}}]{Tanihata1985}%
  \BibitemOpen
  \bibfield  {author} {\bibinfo {author} {\bibfnamefont {I.}~\bibnamefont
  {Tanihata}}, \bibinfo {author} {\bibfnamefont {H.}~\bibnamefont {Hamagaki}},
  \bibinfo {author} {\bibfnamefont {O.}~\bibnamefont {Hashimoto}}, \bibinfo
  {author} {\bibfnamefont {Y.}~\bibnamefont {Shida}}, \bibinfo {author}
  {\bibfnamefont {N.}~\bibnamefont {Yoshikawa}}, \bibinfo {author}
  {\bibfnamefont {K.}~\bibnamefont {Sugimoto}}, \bibinfo {author}
  {\bibfnamefont {O.}~\bibnamefont {Yamakawa}}, \bibinfo {author}
  {\bibfnamefont {T.}~\bibnamefont {Kobayashi}},\ and\ \bibinfo {author}
  {\bibfnamefont {N.}~\bibnamefont {Takahashi}},\ }\bibfield  {title} {\bibinfo
  {title} {Measurements of interaction cross sections and nuclear radii in the
  light $p$-shell region},\ }\href
  {https://doi.org/10.1103/PhysRevLett.55.2676} {\bibfield  {journal} {\bibinfo
   {journal} {Phys. Rev. Lett.}\ }\textbf {\bibinfo {volume} {55}},\ \bibinfo
  {pages} {2676} (\bibinfo {year} {1985})}\BibitemShut {NoStop}%
\bibitem [{\citenamefont {Austern}\ \emph {et~al.}(1987)\citenamefont
  {Austern}, \citenamefont {Iseri}, \citenamefont {Kamimura}, \citenamefont
  {Kawai}, \citenamefont {Rawitscher},\ and\ \citenamefont
  {Yahiro}}]{Austern1987125}%
  \BibitemOpen
  \bibfield  {author} {\bibinfo {author} {\bibfnamefont {N.}~\bibnamefont
  {Austern}}, \bibinfo {author} {\bibfnamefont {Y.}~\bibnamefont {Iseri}},
  \bibinfo {author} {\bibfnamefont {M.}~\bibnamefont {Kamimura}}, \bibinfo
  {author} {\bibfnamefont {M.}~\bibnamefont {Kawai}}, \bibinfo {author}
  {\bibfnamefont {G.}~\bibnamefont {Rawitscher}},\ and\ \bibinfo {author}
  {\bibfnamefont {M.}~\bibnamefont {Yahiro}},\ }\bibfield  {title} {\bibinfo
  {title} {Continuum-discretized coupled-channels calculations for three-body
  models of deuteron-nucleus reactions},\ }\href
  {https://doi.org/https://doi.org/10.1016/0370-1573(87)90094-9} {\bibfield
  {journal} {\bibinfo  {journal} {Phys. Rep.}\ }\textbf {\bibinfo {volume}
  {154}},\ \bibinfo {pages} {125} (\bibinfo {year} {1987})}\BibitemShut
  {NoStop}%
\bibitem [{\citenamefont {Thompson}(1988)}]{Thompson1988167}%
  \BibitemOpen
  \bibfield  {author} {\bibinfo {author} {\bibfnamefont {I.~J.}\ \bibnamefont
  {Thompson}},\ }\bibfield  {title} {\bibinfo {title} {Coupled reaction
  channels calculations in nuclear physics},\ }\href
  {https://doi.org/https://doi.org/10.1016/0167-7977(88)90005-6} {\bibfield
  {journal} {\bibinfo  {journal} {Comput. Phys. Rep.}\ }\textbf {\bibinfo
  {volume} {7}},\ \bibinfo {pages} {167} (\bibinfo {year} {1988})}\BibitemShut
  {NoStop}%
\bibitem [{\citenamefont {Rhoades-Brown}\ and\ \citenamefont
  {Braun-Munzinger}(1984)}]{Rhoadesbrown198419}%
  \BibitemOpen
  \bibfield  {author} {\bibinfo {author} {\bibfnamefont {M.}~\bibnamefont
  {Rhoades-Brown}}\ and\ \bibinfo {author} {\bibfnamefont {P.}~\bibnamefont
  {Braun-Munzinger}},\ }\bibfield  {title} {\bibinfo {title} {Explanation of
  sub-barrier fusion enhancement in a coupled channels model},\ }\href
  {https://doi.org/https://doi.org/10.1016/0370-2693(84)92047-1} {\bibfield
  {journal} {\bibinfo  {journal} {Physics Letters B}\ }\textbf {\bibinfo
  {volume} {136}},\ \bibinfo {pages} {19} (\bibinfo {year} {1984})}\BibitemShut
  {NoStop}%
\bibitem [{\citenamefont {Chamon}\ \emph {et~al.}(2021)\citenamefont {Chamon},
  \citenamefont {Carlson},\ and\ \citenamefont {Gasques}}]{Chamon2021108061}%
  \BibitemOpen
  \bibfield  {author} {\bibinfo {author} {\bibfnamefont {L.}~\bibnamefont
  {Chamon}}, \bibinfo {author} {\bibfnamefont {B.}~\bibnamefont {Carlson}},\
  and\ \bibinfo {author} {\bibfnamefont {L.}~\bibnamefont {Gasques}},\
  }\bibfield  {title} {\bibinfo {title} {{São Paulo potential version 2 (SPP2)
  and Brazilian nuclear potential (BNP)}},\ }\href
  {https://doi.org/https://doi.org/10.1016/j.cpc.2021.108061} {\bibfield
  {journal} {\bibinfo  {journal} {Computer Physics Communications}\ }\textbf
  {\bibinfo {volume} {267}},\ \bibinfo {pages} {108061} (\bibinfo {year}
  {2021})}\BibitemShut {NoStop}%
\bibitem [{\citenamefont {Carlson}\ and\ \citenamefont
  {Hirata}(2000)}]{Carlson2000}%
  \BibitemOpen
  \bibfield  {author} {\bibinfo {author} {\bibfnamefont {B.~V.}\ \bibnamefont
  {Carlson}}\ and\ \bibinfo {author} {\bibfnamefont {D.}~\bibnamefont
  {Hirata}},\ }\bibfield  {title} {\bibinfo {title} {Dirac-hartree-bogoliubov
  approximation for finite nuclei},\ }\href
  {https://doi.org/10.1103/PhysRevC.62.054310} {\bibfield  {journal} {\bibinfo
  {journal} {Phys. Rev. C}\ }\textbf {\bibinfo {volume} {62}},\ \bibinfo
  {pages} {054310} (\bibinfo {year} {2000})}\BibitemShut {NoStop}%
\bibitem [{\citenamefont {Druet}\ \emph {et~al.}(2010)\citenamefont {Druet},
  \citenamefont {Baye}, \citenamefont {Descouvemont},\ and\ \citenamefont
  {Sparenberg}}]{Druet201088}%
  \BibitemOpen
  \bibfield  {author} {\bibinfo {author} {\bibfnamefont {T.}~\bibnamefont
  {Druet}}, \bibinfo {author} {\bibfnamefont {D.}~\bibnamefont {Baye}},
  \bibinfo {author} {\bibfnamefont {P.}~\bibnamefont {Descouvemont}},\ and\
  \bibinfo {author} {\bibfnamefont {J.-M.}\ \bibnamefont {Sparenberg}},\
  }\bibfield  {title} {\bibinfo {title} {{CDCC calculations with the
  Lagrange-mesh technique}},\ }\href
  {https://doi.org/https://doi.org/10.1016/j.nuclphysa.2010.05.060} {\bibfield
  {journal} {\bibinfo  {journal} {Nuclear Physics A}\ }\textbf {\bibinfo
  {volume} {845}},\ \bibinfo {pages} {88} (\bibinfo {year} {2010})}\BibitemShut
  {NoStop}%
\bibitem [{\citenamefont {Druet}\ and\ \citenamefont
  {Descouvemont}(2012)}]{Druet2012}%
  \BibitemOpen
  \bibfield  {author} {\bibinfo {author} {\bibfnamefont {T.}~\bibnamefont
  {Druet}}\ and\ \bibinfo {author} {\bibfnamefont {P.}~\bibnamefont
  {Descouvemont}},\ }\bibfield  {title} {\bibinfo {title} {{Continuum Effects
  in the Scattering of Exotic Nuclei}},\ }\href
  {https://doi.org/10.1140/epja/i2012-12147-9} {\bibfield  {journal} {\bibinfo
  {journal} {Eur. Phys. J. A}\ }\textbf {\bibinfo {volume} {48}},\ \bibinfo
  {pages} {147} (\bibinfo {year} {2012})}\BibitemShut {NoStop}%
\bibitem [{\citenamefont {Baye}(2015)}]{Baye20151}%
  \BibitemOpen
  \bibfield  {author} {\bibinfo {author} {\bibfnamefont {D.}~\bibnamefont
  {Baye}},\ }\bibfield  {title} {\bibinfo {title} {The lagrange-mesh method},\
  }\href {https://doi.org/https://doi.org/10.1016/j.physrep.2014.11.006}
  {\bibfield  {journal} {\bibinfo  {journal} {Phys. Rep.}\ }\textbf {\bibinfo
  {volume} {565}},\ \bibinfo {pages} {1} (\bibinfo {year} {2015})},\ \bibinfo
  {note} {the Lagrange-mesh method}\BibitemShut {NoStop}%
\bibitem [{\citenamefont {Shaikh}\ \emph {et~al.}(2014)\citenamefont {Shaikh},
  \citenamefont {Roy}, \citenamefont {Rajbanshi}, \citenamefont {Pradhan},
  \citenamefont {Mukherjee}, \citenamefont {Basu}, \citenamefont {Pal},
  \citenamefont {Nanal}, \citenamefont {Pillay},\ and\ \citenamefont
  {Shrivastava}}]{Shaikh2014}%
  \BibitemOpen
  \bibfield  {author} {\bibinfo {author} {\bibfnamefont {M.~M.}\ \bibnamefont
  {Shaikh}}, \bibinfo {author} {\bibfnamefont {S.}~\bibnamefont {Roy}},
  \bibinfo {author} {\bibfnamefont {S.}~\bibnamefont {Rajbanshi}}, \bibinfo
  {author} {\bibfnamefont {M.~K.}\ \bibnamefont {Pradhan}}, \bibinfo {author}
  {\bibfnamefont {A.}~\bibnamefont {Mukherjee}}, \bibinfo {author}
  {\bibfnamefont {P.}~\bibnamefont {Basu}}, \bibinfo {author} {\bibfnamefont
  {S.}~\bibnamefont {Pal}}, \bibinfo {author} {\bibfnamefont {V.}~\bibnamefont
  {Nanal}}, \bibinfo {author} {\bibfnamefont {R.~G.}\ \bibnamefont {Pillay}},\
  and\ \bibinfo {author} {\bibfnamefont {A.}~\bibnamefont {Shrivastava}},\
  }\bibfield  {title} {\bibinfo {title} {Investigation of $^{6}\mathrm{Li}$ +
  $^{64}\mathrm{Ni}$ fusion at near-barrier energies},\ }\href
  {https://doi.org/10.1103/PhysRevC.90.024615} {\bibfield  {journal} {\bibinfo
  {journal} {Phys. Rev. C}\ }\textbf {\bibinfo {volume} {90}},\ \bibinfo
  {pages} {024615} (\bibinfo {year} {2014})}\BibitemShut {NoStop}%
\bibitem [{\citenamefont {Pakou}\ \emph {et~al.}(2009)\citenamefont {Pakou},
  \citenamefont {Rusek}, \citenamefont {Alamanos}, \citenamefont {Aslanoglou},
  \citenamefont {Kokkoris}, \citenamefont {Lagoyannis}, \citenamefont
  {Mertzimekis}, \citenamefont {Musumarra}, \citenamefont {Nicolis},
  \citenamefont {Pierroutsakou},\ and\ \citenamefont {Roubos}}]{Pakou2009}%
  \BibitemOpen
  \bibfield  {author} {\bibinfo {author} {\bibfnamefont {A.}~\bibnamefont
  {Pakou}}, \bibinfo {author} {\bibfnamefont {K.}~\bibnamefont {Rusek}},
  \bibinfo {author} {\bibfnamefont {N.}~\bibnamefont {Alamanos}}, \bibinfo
  {author} {\bibfnamefont {X.}~\bibnamefont {Aslanoglou}}, \bibinfo {author}
  {\bibfnamefont {M.}~\bibnamefont {Kokkoris}}, \bibinfo {author}
  {\bibfnamefont {A.}~\bibnamefont {Lagoyannis}}, \bibinfo {author}
  {\bibfnamefont {T.~J.}\ \bibnamefont {Mertzimekis}}, \bibinfo {author}
  {\bibfnamefont {A.}~\bibnamefont {Musumarra}}, \bibinfo {author}
  {\bibfnamefont {N.~G.}\ \bibnamefont {Nicolis}}, \bibinfo {author}
  {\bibfnamefont {D.}~\bibnamefont {Pierroutsakou}},\ and\ \bibinfo {author}
  {\bibfnamefont {D.}~\bibnamefont {Roubos}},\ }\bibfield  {title} {\bibinfo
  {title} {Total reaction and fusion cross sections at sub- and near-barrier
  energies for the system $^{7}\mathrm{Li}$ + $^{28}\mathrm{Si}$},\ }\href
  {https://doi.org/10.1140/epja/i2008-10702-7} {\bibfield  {journal} {\bibinfo
  {journal} {Eur. Phys. J. A}\ }\textbf {\bibinfo {volume} {39}},\ \bibinfo
  {pages} {187} (\bibinfo {year} {2009})}\BibitemShut {NoStop}%
\bibitem [{\citenamefont {Sinha}\ \emph {et~al.}(2010)\citenamefont {Sinha},
  \citenamefont {Majumdar}, \citenamefont {Basu}, \citenamefont {Roy},
  \citenamefont {Bhattacharya}, \citenamefont {Biswas}, \citenamefont
  {Pradhan}, \citenamefont {Palit}, \citenamefont {Mazumdar},\ and\
  \citenamefont {Kailas}}]{Sinha2010}%
  \BibitemOpen
  \bibfield  {author} {\bibinfo {author} {\bibfnamefont {M.}~\bibnamefont
  {Sinha}}, \bibinfo {author} {\bibfnamefont {H.}~\bibnamefont {Majumdar}},
  \bibinfo {author} {\bibfnamefont {P.}~\bibnamefont {Basu}}, \bibinfo {author}
  {\bibfnamefont {S.}~\bibnamefont {Roy}}, \bibinfo {author} {\bibfnamefont
  {R.}~\bibnamefont {Bhattacharya}}, \bibinfo {author} {\bibfnamefont
  {M.}~\bibnamefont {Biswas}}, \bibinfo {author} {\bibfnamefont {M.~K.}\
  \bibnamefont {Pradhan}}, \bibinfo {author} {\bibfnamefont {R.}~\bibnamefont
  {Palit}}, \bibinfo {author} {\bibfnamefont {I.}~\bibnamefont {Mazumdar}},\
  and\ \bibinfo {author} {\bibfnamefont {S.}~\bibnamefont {Kailas}},\
  }\bibfield  {title} {\bibinfo {title} {Sub- and above-barrier fusion of
  loosely bound $^{6}\mathrm{Li}$ with $^{28}\mathrm{Si}$},\ }\href
  {https://doi.org/10.1140/epja/i2010-10976-0} {\bibfield  {journal} {\bibinfo
  {journal} {Eur. Phys. J. A}\ }\textbf {\bibinfo {volume} {44}},\ \bibinfo
  {pages} {403} (\bibinfo {year} {2010})}\BibitemShut {NoStop}%
\bibitem [{\citenamefont {Hugi}\ \emph {et~al.}(1981)\citenamefont {Hugi},
  \citenamefont {Lang}, \citenamefont {M{\"u}ller}, \citenamefont {Ungricht},
  \citenamefont {Bodek}, \citenamefont {Jarczyk}, \citenamefont {Kamys},
  \citenamefont {Magiera}, \citenamefont {Strza{\l}kowski},\ and\ \citenamefont
  {Willim}}]{Hugi1981}%
  \BibitemOpen
  \bibfield  {author} {\bibinfo {author} {\bibfnamefont {M.}~\bibnamefont
  {Hugi}}, \bibinfo {author} {\bibfnamefont {J.}~\bibnamefont {Lang}}, \bibinfo
  {author} {\bibfnamefont {R.}~\bibnamefont {M{\"u}ller}}, \bibinfo {author}
  {\bibfnamefont {E.}~\bibnamefont {Ungricht}}, \bibinfo {author}
  {\bibfnamefont {K.}~\bibnamefont {Bodek}}, \bibinfo {author} {\bibfnamefont
  {L.}~\bibnamefont {Jarczyk}}, \bibinfo {author} {\bibfnamefont
  {B.}~\bibnamefont {Kamys}}, \bibinfo {author} {\bibfnamefont
  {A.}~\bibnamefont {Magiera}}, \bibinfo {author} {\bibfnamefont
  {A.}~\bibnamefont {Strza{\l}kowski}},\ and\ \bibinfo {author} {\bibfnamefont
  {G.}~\bibnamefont {Willim}},\ }\bibfield  {title} {\bibinfo {title} {Fusion
  and direct reactions for strongly and weakly bound projectiles},\ }\href
  {https://doi.org/10.1016/0375-9474(81)90739-9} {\bibfield  {journal}
  {\bibinfo  {journal} {Nuclear Physics A}\ }\textbf {\bibinfo {volume}
  {368}},\ \bibinfo {pages} {173} (\bibinfo {year} {1981})}\BibitemShut
  {NoStop}%
\bibitem [{\citenamefont {Rath}\ \emph
  {et~al.}(2009{\natexlab{b}})\citenamefont {Rath}, \citenamefont {Santra},
  \citenamefont {Mahata}, \citenamefont {Tripathi}, \citenamefont {Parkar},
  \citenamefont {Palit},\ and\ \citenamefont {Singh}}]{Rath2009a}%
  \BibitemOpen
  \bibfield  {author} {\bibinfo {author} {\bibfnamefont {P.~K.}\ \bibnamefont
  {Rath}}, \bibinfo {author} {\bibfnamefont {S.}~\bibnamefont {Santra}},
  \bibinfo {author} {\bibfnamefont {K.}~\bibnamefont {Mahata}}, \bibinfo
  {author} {\bibfnamefont {R.}~\bibnamefont {Tripathi}}, \bibinfo {author}
  {\bibfnamefont {V.~V.}\ \bibnamefont {Parkar}}, \bibinfo {author}
  {\bibfnamefont {R.}~\bibnamefont {Palit}},\ and\ \bibinfo {author}
  {\bibfnamefont {N.~L.}\ \bibnamefont {Singh}},\ }\bibfield  {title} {\bibinfo
  {title} {Complete versus incomplete fusion in
  $^{6}\mathrm{Li}$+$^{144}\mathrm{Sm}$ reaction},\ }\href@noop {} {\bibfield
  {journal} {\bibinfo  {journal} {Proceedings of the International Symposium on
  Nuclear Physics}\ ,\ \bibinfo {pages} {310}} (\bibinfo {year}
  {2009}{\natexlab{b}})}\BibitemShut {NoStop}%
\end{thebibliography}%

\end{document}